\documentclass{llncs}
\usepackage{latexsym,subfigure}
\usepackage{amsfonts}
\usepackage{amsmath,paralist}
\usepackage{times}
\usepackage{graphics,color}
\usepackage{url}
\usepackage{graphicx}
\usepackage{wrapfig}
\usepackage{url}

\begin{document}

\title{Cybersecurity Dynamics: A Foundation for the Science of Cybersecurity}

\author{Shouhuai Xu}

\institute{Laboratory for Cybersecurity Dynamics, Department of Computer Science,
University of Texas at San Antonio.
Email: \email{shxu@cs.utsa.edu}. Web: \url{www.cs.utsa.edu/~shxu}}

\maketitle

\abstract{Cybersecurity Dynamics is new concept that aims to achieve the modeling, analysis, quantification, and management of cybersecurity from a holistic perspective, rather than from a building-blocks perspective. It is centered at modeling and analyzing the attack-defense interactions in cyberspace, which cause a ``natural'' phenomenon --- the evolution of the global cybersecurity state. In this Chapter, we systematically introduce and review the Cybersecurity Dynamics foundation for the Science of Cybersecurity. We review the core concepts, technical approaches, research axes, and results that have been obtained in this endeavor. We outline a research roadmap towards the ultimate research goal, including a systematic set of technical barriers.}

\section{Introduction}

The fundamental concepts of confidentiality, integrity, and availability have been at the core of information security research over the past decades.
These concepts have led to the development of many building-block techniques, such as cryptographic mechanisms, which can be rigorously analyzed in a sound scientific framework.
This motivated us to seek fundamental concepts and frameworks that can guide our investigation of cybersecurity, which has to be understood from a holistic perspective (i.e., by treating a network of interest as a whole, rather than investigating their building-blocks separately).

In the course of our endeavor, the concept of {\em cybersecurity dynamics} emerges \cite{XuCD}.
Intuitively, the concept of cybersecurity dynamics reflects the {\em evolution} of the global cybersecurity state of a network, where ``evolution'' is caused by the interactions between the human parties involved --- dubbed {\em attack-defense interactions}. The human parties involved include attackers who wage attacks against a network, defenders who employ defense mechanisms to protect a network in question, and users who may be exploited by the attackers to wage attacks.

The concept of Cybersecurity Dynamics is appealing because of the following.
First, the global cybersecurity state of a network reflects the real-time situation, which ``naturally'' evolves over time because of the attack-defense interactions.
Knowing the real-time global cybersecurity state or situation is of high interest to cyber defense decision-makers, who often need to adjust their defense posture (including policies, architectures, and mechanisms) to mitigate or minimize the damage of cyber attacks. Second, the effects of employing new cyber defense postures are reflected by the resulting global cybersecurity state. This means that we can compare the effectiveness of one defense posture against another.
Third, looking at the evolution of the global cybersecurity state allows us to build systematic models with {\em descriptive} power (i.e., characterizing what phenomenon can happen under what circumstances), {\em prescriptive} power (i.e., guiding the adjustment to defense postures to mitigate or minimize the damage of cyber attacks), and {\em predictive} power (i.e., forecasting what will happen with or without making adjustments to the defense posture). Four, modeling the evolution of the global cybersecurity state makes security quantification an inherent task, which paves the way for quantitative decision-making in the course of cyber defense operations. In particular, the concept of Cybersecurity Dynamics naturally leads to the notion of {\em macroscopic cybersecurity}, with models that will use parameters to describe or represent (among other things) attacks and defenses.

\smallskip

\noindent{\bf Our contributions}. The present chapter systematically refines and extends an earlier treatment of the Cybersecurity Dynamics foundation given in \cite{XuCybersecurityDynamicsHotSoS2014}, while accommodating the many advancements that have been made during the past few years. More specifically, we systematically introduce and review the Cybersecurity Dynamics foundation (or framework), while focusing on three orthogonal, coherent ``axes'': (i) the cybersecurity metrics axis aims to develop a systematic set of metrics that can adequately describe cybersecurity; (ii) the cybersecurity first-principle modeling and analysis axis aims to establish cybersecurity laws governing the evolution of the global cybersecurity state; and (iii) the cybersecurity data analytics axis aims to extract model parameters and validate/invalidate models developed in the first-principle modeling and analysis axis. In particular, we discuss the deep connections between these three axes. Despite the many efforts and significant results, there are many outstanding problems that have yet to be tackled. We hope the present chapter will inspire many more studies to address the many open problems.

\smallskip

\noindent{\bf Chapter outline}.
The chapter is organized as follows.
Section \ref{sec:overview} presents an overview of the Cybersecurity Dynamics foundation.
Section \ref{sec:metrics} reviews the recent advancement in cybersecurity metrics research.
Section \ref{sec:first-principle} reviews the recent advancement in cybersecurity first-principle modeling and analysis.
Section \ref{sec:data-analytics} reviews the recent advancement in cybersecurity data analytics.
Section \ref{sec:future-research} discusses future research directions, including technical barriers that need to be tackled.
Section \ref{sec:related-work} reviews related prior studies.
Section \ref{sec:conclusion} concludes the present chapter.

\section{Overview of the Cybersecurity Dynamics Foundation}
\label{sec:overview}

\subsection{Terminology}

By ``network'' we mean an arbitrary (cyber, cyber-physical, Internet of Things or IoT) network of interest that is enabled or interconnected by the TCP/IP technology, regardless of the underlying communication being wired or wireless.
A network can have an arbitrarily large size  (e.g., an enterprise network or even the entire cyberspace).
By ``computer'' we mean a computer or device (e.g., smartphones, IoT devices) with a software stack, which typically includes some applications, library functions, and an operating system.

A network is protected by some {\em defenders}, who may or may not be under the same administrative jurisdiction (e.g., a network of interest consisting of multiple independently managed enterprise networks). Each network has a number of {\em users}, who are often subject to attacks (e.g., social-engineering attacks).
The {\em attacker} attempts to compromise the computers in a network, by exploiting weaknesses in the network software and hardware as well as weaknesses in the users or defenders (e.g., making them become insider threats).

In the context of the present chapter, the terms {\em cybersecurity} and {\em security} are used interchangeably.
In order to model cybersecurity from a holistic perspective (in contrast to building-block perspectives), we need to have the notion of {\em model resolution}, reflecting the level of abstraction. For example, we can treat a computer or software component as an indivisible unit, dubbed ``{\em atoms}'' of a model. Throughout the chapter, we will use the term ``atom'' to indicate the unit from a modeling point of view. Because each ``atom'' will be represented as a vertex or node in a graph-theoretic model, we also call an ``atom'' a {\em node}.
When we treat a computer as a unit or ``atom'', we are dealing with a coarse-grained model because the internal components of the computer are treated as transparent. As a consequence, compromise of any program in the user space of a computer would force us to treat the entire computer as compromised.
When we treat a software component (e.g., software program or even program function) as an ``atom'', we are dealing with a fine-grained model because the compromise of one component in a computer (e.g., application) does not necessarily mean the compromise of another component in the same computer (e.g., the operating system).

For each ``atom'' mentioned above, we can define its {\em security state}, which can be either {\em secure} but possibly vulnerable to attacks because it contain some vulnerabilities, or {\em compromised}. In the real world, the security state of an ``atom'' is dynamic (i.e., changing over time), rather than static, because it can become compromised (because of some attack actions), then become secure (because of some defense actions), then become compromised, and so on. This naturally leads to the view that the security state evolves.  We call the security state of an ``atom'' a {\em local} cybersecurity state because it deals with an individual ``atom''; we call the security state of an entire network the {\em global} cybersecurity state, which can be represented as a vector of the local cybersecurity states of the ``atoms''.

\begin{figure}[htbp!]
\centering
\includegraphics[width=.8\textwidth]{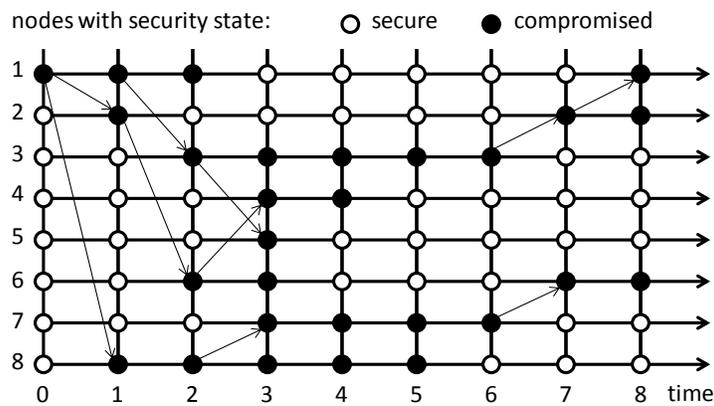}
\caption{Illustration of the evolution of the global cybersecurity state in a small network of 8 ``atoms'' at an appropriate model resolution. The ``atoms'' are represented as {\em nodes} (e.g., computers, devices, or software components). In the discrete-time model, each ``atom'' or node has a cybersecurity state at any point in time, either {\em secure} (represented as an empty circle) or {\em compromised} (represented as a filled circle) in this example. Each arrow represents a successful attack from a compromised node against a secure node, causing the latter to become compromised. A compromised node may become secure again because of some defense activities. A secure node may be attacked by multiple compromised nodes at the same time.}
\label{fig:state-evolution}
\end{figure}

Figure \ref{fig:state-evolution} illustrates the evolution of the global cybersecurity state of a network, reflected by the evolution of the local cybersecurity states of individual ``atoms'' that are represented as ``nodes'' 1, $\ldots$, 8. In this illustration, a node has two possible states at any point in time, {\em secure} (empty circle) or {\em compromised} (filled circle).
A secure node may be attacked by one or multiple compromised nodes and then become compromised; a compromised node may become secure again because of some defense activities.
An arrow indicates a successful attack.

\subsection{Research Objectives}

The evolution of the global cybersecurity state, as illustrated in Figure \ref{fig:state-evolution}, is a {\em natural} phenomenon in cyberspace. The core research objectives of Cybersecurity Dynamics are centered at {\em understanding}, {\em managing} (or controlling), and {\em forecasting} the evolution. Understanding the evolution means we want to gain deep insights into the laws that govern the evolution. For this purpose, we need to build {\em descriptive} models to analyze how the attack-defense interactions govern the evolution of the global cybersecurity state.
Managing the evolution means that we want to mitigate or control, if not minimize, the damage so as to benefit the defender.
For this purpose, we need to build {\em prescriptive} models that can guide the orchestration of cyber defense activities in an optimal or cost-effective fashion.
Forecasting means that we want to be able to forecast or predict the evolution so as to facilitate adaptive and/or proactive cyber defense.
For this purpose, we need to build {\em predictive} models that can forecast, among other things, the evolution of the global cybersecurity state and the incoming threats against a network of interest.

\begin{figure}[htbp!]
\centering
\includegraphics[width=.8\textwidth]{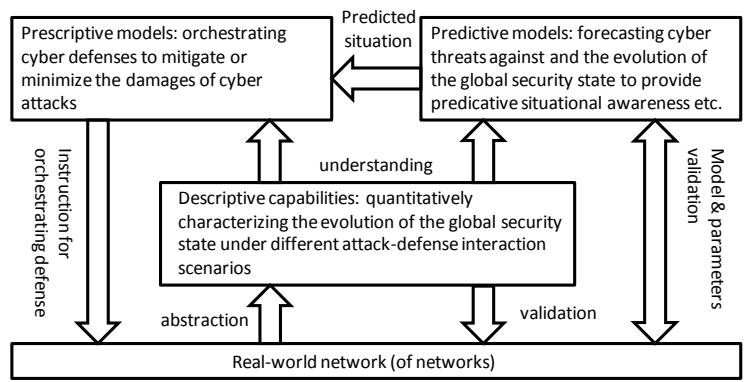}
\caption{Three core research objectives of Cybersecurity Dynamics: descriptive capabilities, prescriptive capabilities, and predictive capabilities.}
\label{fig:research-objectives}
\end{figure}

Figure \ref{fig:research-objectives} highlights the aforementioned three core research objectives and the relationship between them.
Descriptive models are abstracted from the real-world networks by faithfully representing the attack-defense interactions. These models will be validated (or invalidated) according to real-world data or experiments. Predictive models are built on top of the description models and are also validated (or invalidated) according to real-world data.
Prescriptive models are also built on top of descriptive models, while possibly taking into consideration the situations predicted or forecasted by the predictive models.
The prescriptive models will guide the orchestration of cyber defense so as to benefit the defender in a cost-effective, if not optimal, fashion.

\subsection{Scope}

\begin{figure}[htbp!]
\centering
\includegraphics[width=.7\textwidth]{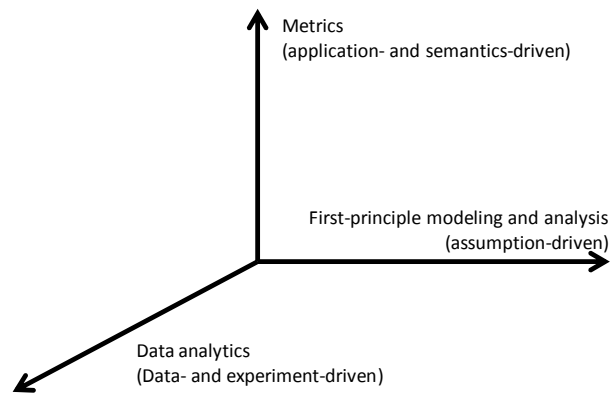}
\caption{Scope of the present chapter: Three research axes towards achieving the research objectives of Cybersecurity Dynamics.}
\label{fig:coordinate-system}
\end{figure}

Figure \ref{fig:coordinate-system} highlights the scope of the present chapter, which focuses on discussing three axes of Cybersecurity Dynamics research:
(i) Cybersecurity metrics, which are driven by applications (e.g., for orchestrating cyber defenses to mitigate or minimize the damage of cyber attacks)
and semantics (e.g., what aspects of cybersecurity would reflect the competence of cyber defense?).
(ii) Cybersecurity first-principle modeling and analysis, which are driven by assumptions. First-principle models are useful in the absence of real-world data and can be inspired by the properties exhibited by real-world datasets.
(iii) Cybersecurity data analytics, which are driven by real-world data or experiments.

\begin{figure}[htbp!]
\centering
\includegraphics[width=.8\textwidth]{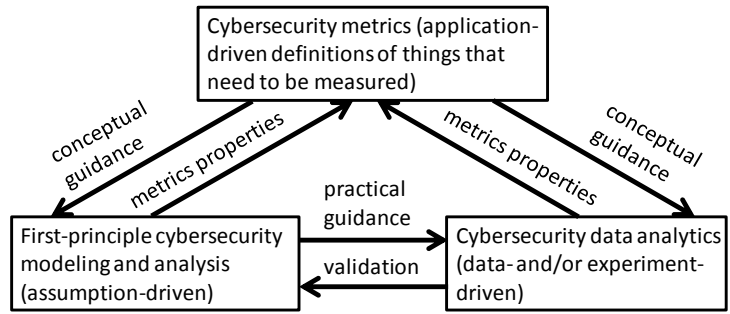}
\caption{Relationship between the three research axes.}
\label{fig:three-axis-relationship}
\end{figure}

Figure \ref{fig:three-axis-relationship} highlights the relationship between the three research axes.
The cybersecurity metrics axis aims to rigorously define metrics to measure and quantify cybersecurity from a holistic perspective, and therefore provides conceptual guidance to the other two axes because those quantitative models are often centered at some metrics. Along this axis, significant progress has been made \cite{Pendleton16,8017389,Noel2017,Cho16-milcom,XuHotSoS2018Firewall,XuHotSoS2018Diversity,XuAgiliy2018manuscript,XuSTRAM2018manuscript}.

The cybersecurity first-principle modeling and analysis axis aims to build, under appropriate assumptions, mathematical models to describe the evolution of the global cybersecurity state caused by cyber attack-defense interactions. By ``first-principle'' we mean the use of as-simple-as-possible models with as-few-as-possible parameters, while making as-weak-as-possible assumptions; of course, these models must make sense from a cybersecurity perspective and can be validated/invalidated (e.g., through the validation/invalidation of the assumptions they make). This axis aims to establish cybersecurity laws governing the evolution of the global cybersecurity state. For example, these first-principle models aim to derive macroscopic phenomena (or characteristics or properties) from the underlying microscopic attack-defense interactions. This axis supports the cybersecurity metrics axis by providing insights into the properties of metrics (e.g., do they converge or oscillate over time), and provides practical guidance to the cybersecurity data analytics axis (e.g., by showing that some model parameters are necessary and therefore cannot be replaced with any alternatives). Along this axis, significant progress has been made \cite{XuTDSC2011,XuTAAS2012,XuTDSC2012,XuInternetMath2012,XuGameSec13,XuTAAS2014,XuQuantitativeSecurityHotSoS2014,XuEmergentBehaviorHotSoS2014,XuHotSOS14-MTD,XuInternetMath2015Dependence,XuInternetMath2015ACD,XuHotSoS2015,XuTNSE2018}.

The cybersecurity data analytics axis aims to use data- and/or experiment-driven studies to obtain model parameters and validate/invalidate first-principle models.
This is because first-principle models typically, and legitimately, assume away the obtaining of model parameters.
This axis supports the cybersecurity metrics axis by providing insights into the properties of metrics (e.g., some metrics are hard or costly to measure, suggesting the need to define and use alternate metrics), and helps validate first-principle models (e.g., by showing that an assumption underlying a first-principle model is not valid).
Along this axis, significant progress has been made \cite{XuTIFS2013,XuInTrust2014,XuTIFS2015,XuPLoSOne2015,XuVineCopula2015,XuMarkerPointProcess2017,XuJournalAppliedStatistics2018,XuTIFSDataBreach2018}.

\section{Cybersecurity Metrics}
\label{sec:metrics}

The most outstanding open problem in cybersecurity research is arguably cybersecurity metrics \cite{DBLP:journals/tdsc/NicolST04,Pendleton16,8017389}. Despite its clear importance, the problem is largely open as evidenced by the fact that it has been constantly listed as one of the hard problems \cite{IRC-hardproblemlist,NITRD,NSAHardProblemList}. Recently, the problem has received systematic attention \cite{Pendleton16,8017389,Noel2017,Cho16-milcom,XuHotSoS2018Firewall,XuHotSoS2018Diversity,XuAgiliy2018manuscript,XuSTRAM2018manuscript}.

In Cybersecurity Dynamics \cite{XuCybersecurityDynamicsHotSoS2014,Pendleton16}, the following five kinds of cybersecurity metrics have been proposed to systematically describe the evolution of the global cybersecurity state \cite{Pendleton16}: (i) metrics for describing a network including its configurations; (ii) metrics for describing systems and human vulnerabilities; (iii) metrics for describing defenses employed to protect networks; (iv) metrics for describing cyber attacks (i.e., threat models); and (v) metrics for describing the global cybersecurity state or cybersecurity situational awareness.

Specifically,  let $security\_state(t)$ denote the global cybersecurity state at time $t$, $C(t)$ denote a network of interest at time $t$ (including its hardware and software configurations),
$L(t)$ denote the vulnerabilities in the network at time $t$ (including possibly zero-day vulnerabilities, human factors with uncertainty), $D(t)$ denote the defense posture at time $t$ (i.e., the defense that are employed at time $t$ to protect the network), and $A(t)$ denote the attacks that are waged against the network at time $t$. The framework aims to obtain families of mathematical functions, denoted by $\{f\}$, such that
\begin{equation}
\label{eq:overall-equation}
security\_state(t)= f(C(t), L(t), D(t), A(t)).
\end{equation}
Eq. \eqref{eq:overall-equation}, once achieved, has many applications. For example, it allows us to compare the global cybersecurity of networks deploying two different configurations, say $C(t)$ vs. $C'(t)$, or two different defense postures, say $D(t)$ vs. $D'(t)$, through the difference between the corresponding evolution of $security\_state(t)$ and $security\_state'(t)$ over time.
As we will discuss later, some concrete $f$'s have been investigated in the cybersecurity first-principle modeling and analysis axis and the cybersecurity data analytics axis.

In what follows, we discuss how to obtain mathematical representations of network configurations $C(t)$, vulnerabilities $L(t)$, defense postures $D(t)$, and threats $A(t)$.
These representations naturally lead to quantitative metrics.

\subsection{Representation of Network Configuration and Metrics}

\noindent{\bf Representation}. At a high level, configurations can be reflected by an {\em attack-defense structure}, which can be described as a graph $G(t)=(V(t),E(t))$, where $V(t)$ is the node or vertex set at time $t$, and $E(t)$ is the edge or arc set at time $t$. A node $v\in V(t)$ represents an ``atom'' mentioned above (e.g., a computer or software component). An edge or arc $(u,v)\in E(t)$ means that node $u$ can attack node $v$, meaning that the communication from node $u$ to node $v$ may not be filtered, for example, by host-based intrusion prevention (when $u$ and $v$ belong to the same computer) or by network-based intrusion prevention (when $u$ and $v$ represent, or belong to, different computers). Moreover, $(u,v)\in E(t)$ means that the compromise of node $u$ can cause the compromise of node $v$. Note that $E(t)$ does not necessarily represent the physical network topology in general (except perhaps for sensor networks or IoT networks where nodes can only afford to have short-range communications); in general, $(u,v)\in E(t)$ represents a communication link or path in a network. It turns out that filtering unauthorized communication relations $(u,v)\notin E(t)$ is an important defense means (see, for example, \cite{XuTAAS2012,XuTNSE2018,XuHotSoS2018Firewall,XuHotSoS2018Diversity}).

Recently, researchers have started to investigate how to represent networks at finer granularities \cite{XuHotSoS2018Firewall,XuHotSoS2018Diversity}.
Suppose a network of interest is composed of $n(t)$ computers or devices at time $t$.
In order to obtain the attack-defense structure $G(t)=(V(t),E(t))$, we need to first represent the {\em software stacks} on each computer or device, meaning that we need to model the applications, operating systems, and possibly library functions. Then, a computer or device, denoted by $i$, may be represent by a graph $G_i(t)=(V_i(t),E_i(t))$, where $v\in V_i$ represents an ``atom'' (e.g., application, operating system, or function), and $(u,v)\in E_i(t)$ means either $u$ can call $v$ (i.e., caller-callee dependence relation) or $u$ can communicate with $v$ (i.e., inter-application communication relation). Another edge set $E_0(t)$ may be defined to represent the authorized inter-computer communications within the network at time $t$.
Yet another edge $E_*(t)$ may be defined to represent the authorized inter-network communication relations between the network and the external networks (i.e., internal-external communication relations). Note that $E_i(t)$ reflects a host-based access control policy (if employed), and $E_0(t)$ and $E_*(t)$ reflect network-wide access control policies (if employed).
As a result, the attack-defense structure $G(t)=(V(t),E(t))$ may be derived as follows  \cite{XuHotSoS2018Firewall,XuHotSoS2018Diversity}:
\begin{equation*}
V(t)=V_1(t) \cup \ldots \cup V_{n(t)}(t)~~\text{and}~~E(t)=E_1(t) \cup \ldots \cup E_{n(t)}(t) \cup E_0(t) \cup E_*(t).
\end{equation*}

\noindent{\bf Metrics}. Having obtained the graph-theoretic representation $G(t)=(V(t),E(t))$, we may define metrics to characterize $G(t)$. For example, we may use nodes' degree distribution to characterize the structure of $G(t)$; we may characterize the evolution of $G(t)$ over time; we may quantify the difference of two defense policies by comparing the attack-defense structures resulting from their respective employments.

\subsection{Representation of Vulnerabilities and Metrics}

\noindent{\bf Representation}. We propose classifying vulnerabilities into three kinds: software, hardware, and human vulnerabilities, which are all used in a broad sense.
\begin{itemize}
\item We use the term ``software vulnerabilities'' to describe the vulnerabilities in the entire software stack, including applications, library functions, and operating systems. Software vulnerabilities are the root cause of many real-world attacks. For example, the problem of {\em vulnerability detection} is an active research topic (see, for example, \cite{li2016vulpecker,DBLP:conf/sp/KimWLO17,vuldeepecker}).

\item We use the term ``hardware vulnerabilities'' to describe the vulnerabilities in the hardware, architecture, and firmware. The number of hardware vulnerabilities is often much smaller than the number of software vulnerabilities, but the damage caused by a hardware vulnerability is often severe because of the wide use of the hardware.
Two recent examples of hardware vulnerabilities are Spectre and Meltdown (see, for example, \cite{DBLP:journals/corr/abs-1802-03802,DBLP:journals/corr/abs-1801-01203}).

\item We use the term ``human vulnerabilities'' to describe the vulnerabilities of the users and administrators, such as vulnerabilities to social-engineering attacks (e.g., phishing) as well as insider threats and the vulnerabilities caused by the use of weak passwords.
\end{itemize}
Each vulnerability may be associated with a set of attributes. For example, a software vulnerability may have the following attributes: (i) the privilege that is required in order to exploit the vulnerability (e.g., local access vs. remote access); (ii) what is the chance that there is a zero-day vulnerability in a software component?
(iii) what is the security consequence of the exploitation of a vulnerability?

\smallskip

\noindent{\bf Metrics}. Corresponding to these vulnerabilities, metrics need to be defined to quantify them. Two approaches have been proposed in the literature to measure software vulnerabilities, {\em coarse-grained} vs. {\em fine-grained}.
\begin{itemize}
\item Fine-grained approach: In this approach, vulnerabilities are considered at fine-grained granularities by separating the vulnerabilities of applications, library functions, and operating systems \cite{XuHotSoS2018Firewall,XuHotSoS2018Diversity}.
\item Coarse-grained approach: In this approach, vulnerabilities are often discussed at an aggregate level.
For example, when treating a computer as an ``atom'', we consider the overall vulnerability of a computer, which can be aggregated from the vulnerabilities in the applications, library functions, and operating systems. This approach has been used in numerous cybersecurity first-principle models (see \cite{XuTAAS2012,XuTNSE2018} and the references therein).
\end{itemize}
Similarly, hardware vulnerabilities may be characterized by, for example, the chance that a vulnerability can be exploited;
human vulnerabilities may be described by the chance that a user or defender is vulnerable to social-engineering attacks.

\subsection{Representation of Defenses and Metrics}

\noindent{\bf Representation}.
There are many kinds of defense mechanisms that need to be represented for modeling purposes, such as firewalls, host-based intrusion prevention/detection systems, and network-based intrusion prevention/detection systems. Moreover, access control policies also need to be represented. For example, a tight access control policy would filter or block any unauthorized communication or function call; in contrast, a loose access control policy would not filter or block any unauthorized communication or function call, which can happen when some ``atoms'' are compromised.
For modeling purposes, we classify defenses into preventive, reactive, proactive, adaptive, and active defenses.
\begin{itemize}
\item Preventive defenses aim to prevent attacks from succeeding or even reaching the target of interest. Mechanisms such as whitelisting, access control, and firewall are examples of preventive defenses.
\item Reactive defenses aim to detect successful attacks and ``clean up'' their damage. Mechanisms such as anti-malware tools are examples of reactive defenses.
\item Adaptive defenses aim to dynamically adjust the defense posture so as to mitigate or disrupt ongoing attacks that have been detected by the defender. Examples include the use of Software-Defined Networking (SDN) technology to change network configurations, or route network traffic through dynamically employed network security tools such as firewalls and intrusion prevention/detection systems. A concrete example for protecting systems with known, but unpatched, vulnerabilities is shown in \cite{XuESORICS2018sub}.
\item Proactive defenses aim to dynamically adjust the defense posture so as to mitigate or disrupt attacks, whose presence is not necessarily known to the defender. Mechanisms such as Moving Target Defense (MTD) are examples of proactive defenses.
\item Active defenses aim to deploy defense mechanisms (or defenseware) to ``patrol'' networks to detect and clean up compromises. In the context of the present Chapter, active defenses are not meant to be ``hacking back'' because the defenseware are deployed within the boundary of the defender's network.
 \end{itemize}

\noindent{\bf Metrics}.
Metrics need to be defined to measure the defense capabilities of a defender. For a preventive defense mechanism, we need to measure what kinds of cyber attacks that can or cannot be prevented by it. For a reactive defense mechanism, its detection capabilities can be measured by the false-positive rate, false-negative rate, and related metrics; similarly, its ``cleaning'' capabilities may not be perfect as well (because there is evidence showing that using multiple anti-malware tools together is not adequate to clean up malware infecting a computer \cite{DBLP:conf/dimva/MohaisenA14,Perdisci:2012:VTF:2420950.2420999,XuASE2012,XuTIFSTrustworthiness2018}).
For adaptive defense, its capabilities before and after an adaptation should be different (e.g., in terms of both attack-prevention and attack-detection capabilities).
For proactive defense, its capabilities can be measured by the extent to which the compromised nodes can be cleaned by such mechanisms.
For active defense, its capabilities can be measured by what kinds of attacks can be detected and cleaned up by such mechanisms.

\subsection{Representation of Attacks and Metrics}

\noindent{\bf Representation}.
There are many kinds of cyber attacks, which can be characterized from multiple perspectives. From the perspective of {\em attack freshness}, which often reflects the {\em attack evasion capability}, we can classify attacks into the following categories:
\begin{itemize}
\item Zero-day attacks: These attacks can be further divided into two sub-categories, depending on the freshness of the vulnerabilities they exploit.
\begin{itemize}
\item Zero-day attacks exploiting zero-day vulnerabilities: These attacks exploit zero-day vulnerabilities which are not known to anyone but the attacker, the exploit writer, or the entity that discovered the vulnerability. These attacks are often difficult to detect, let alone prevent. These attacks can also accommodate the exploitation of newly compromised employees as {\em insider threats}.
\item Zero-day attacks exploiting known vulnerabilities: These attacks exploit known, but unpatched, vulnerabilities, while possibly able to evade any existing defense systems (e.g., intrusion prevention/detection systems).
\end{itemize}
\item Known attacks: These attacks are recognized by defense systems and therefore can be blocked before they cause any damage or detected after they penetrate into computers or devices.
\end{itemize}

From the perspective of {\em attack behaviors}, which often reflect the characteristics of attackers, we can classify attacks into the following categories:
\begin{itemize}
\item Machine-waged attacks: These attacks are largely waged by machines and are largely automated.
\begin{itemize}
\item Push-based attacks: These attacks actively seek to compromise other computers or devices \cite{XuTAAS2012}. Examples of these attacks are computer malware, which actively search for vulnerable victims. Social engineering attacks also fall into this category.
\item Pull-based attacks: These attacks passively wait to compromise other computers or devices \cite{XuTAAS2012}. Examples of these attacks are ``drive-by download'' by which a malicious web server waits for connections from vulnerable browsers and then compromises the latter \cite{ProvosHotbot07}.
\end{itemize}
\item Human-waged attacks: These attacks are largely waged by human attackers and are largely manual.
\begin{itemize}
\item Advanced Persistent Threats (APTs): These attacks are often waged by patient attackers targeting high-value assets. These attacks are often carefully planned.
\item Insider Threats: These attacks are largely waged by compromised users who are authorized with some privileges. These attackers are often victims of social engineering attacks, but are aware of their own malicious activities (in contrast to other victims of social engineering attacks, such as those who are lured to double-click a malicious email attachment or access a malicious website).
\end{itemize}
\end{itemize}

From the perspective of {\em attack objectives}, we can classify attacks into the following categories:
\begin{itemize}
\item Attacks against confidentiality: These attacks attempt to compromise the confidentiality of data, either during transmission, which is possible when the cryptographic protection mechanisms or protocols are flawed, or during storage in computer memory or disks, which is possible by penetrating into the computers \cite{ChowUsenixSecurity04,HarrisonDSN07,XuIntrust09-keysecurity,Guan:2015:PPK:2867539.2867702} or using side-channel attacks \cite{kopf2007information}.
\item Attacks against integrity: These attacks attempt to compromise the integrity of data, either during transmission, which is possible when the cryptographic protection mechanisms or protocols are flawed, or during storage in computer memory or disks, which is possible (for example) when the storage provider is malicious (see, e.g., \cite{DBLP:conf/ccs/JuelsK07,XuCODASPY2011-POR,XuCODASPY2012,XuCCSW12,XuInfocom2014,XuIC2E2015}).
\item Attacks against availability: These attacks attempt to make services unavailable to their users \cite{HussainSIGCOMM03}. These attacks are often waged by many compromised computers or devices, such as botnets \cite{XieNSDI09,DBLP:conf/acsac/DagonGLL07,DBLP:conf/IEEEares/LeonardXS09a,XuACNS2010}.
\end{itemize}
Faithful threat or attack models are important. For example, both random and targeted deletions of nodes from computer networks \cite{AlbertNature00} oversimplifies real-world attacks \cite{XuPRE2011,XuPhysicaA2017}.

\smallskip

\noindent{\bf Metrics}. Many kinds of metrics can be defined to measure attack capabilities, such as (i) the {\em exploits} that can be used by the attacker; (ii) the {\em agility} of the attacker, and (iii) the {\em strategy} that can be used by the attacker.
\begin{itemize}
\item Characterizing exploits: An exploit can be described by its attributes, such as: whether it exploits a zero-day vulnerability or an unpatched but known vulnerability.

\item Characterizing attack agility: This attribute aims to describe how active and agile the attacker is. For example, one attacker may only reactively update its exploits after the defender updates its defenses. The first study at modeling and quantifying the agility of attackers is reported in \cite{XuAgiliy2018manuscript}, which presents a metrics framework for transforming well-defined security metrics (e.g., false-positive rate and false-negative rates) to measure attacker agility.

\item Attack strategies: Examples of attack strategies are Lockheed Martin's Cyber Kill Chain~\cite{CyberKillChainPaper2011} and Mandiant's Attack Life Cycle~\cite{Mandiant}.
A general attack strategy may include the following phases: reconnaissance, weaponization, initial compromise, further reconnaissance, privilege escalation, and
lateral movement. At each phase, metrics need to be defined to measure the attack capabilities.
\end{itemize}

\subsection{Security State Metrics}

For any model resolution (e.g., treating a computer/device as an atom vs. treating a software component as an atom), the security state of an ``atom'' can be in one of multiple states, such as {\em secure} vs. {\em compromised}, denoted by
\begin{eqnarray*}
security\_state(atom,t)=\left\{
\begin{array}{ll}
0 & \text{the ``atom'' is in {\em secure} state at time $t$} \\
1 & \text{the ``atom'' is in {\em compromised} state at time $t$}
\end{array}
\right.
\end{eqnarray*}
Therefore, at any point in time, the {\em global} cybersecurity state can be defined as
$$
global\_security(t)=\frac{\text{the number of ``atoms'' in the {\em compromised} state at time $t$}}{\text{the total number of ``atoms'' at time $t$}},
$$
while noting that the total number of ``atoms'' can dynamically evolve.
This is arguably one of the most fundamental metrics and has been the center of numerous cybersecurity first-principle models \cite{XuCybersecurityDynamicsHotSoS2014}.

\section{Cybersecurity First-Principle Modeling and Analysis}
\label{sec:first-principle}

At a high level, cybersecurity first-principle modeling aims to design and characterize the various kinds of mathematical functions $f$ illustrated in Eq. \eqref{eq:overall-equation}.
Several kinds of $f$'s have been proposed to describe different kinds of attack-defense interactions and the resulting dynamics~\cite{XuCybersecurityDynamicsHotSoS2014}:
preventive and reactive cyber defense dynamics
\cite{XuTDSC2011,XuTAAS2012,XuTDSC2012,XuInternetMath2012,XuGameSec13,XuInternetMath2015Dependence,XuTNSE2018,XuHotSoS2018Firewall,XuHotSoS2018Diversity,XuComplexNetworkSub2018};
adaptive cyber defense dynamics \cite{XuTAAS2014,XuQuantitativeSecurityHotSoS2014};
proactive cyber defense dynamics \cite{XuHotSOS14-MTD}; and
active cyber defense dynamics \cite{XuInternetMath2015ACD,XuHotSoS2015}.

\subsection{Preventive and Reactive Cyber Defense Dynamics}

The systematic preventive and reactive cyber defense dynamics model presented in \cite{XuAINA07} accommodates arbitrary, but time-independent, attack-defense structures $G=(V,E)$, push-based attacks (e.g., malware spreading), and pull-based attacks (e.g., drive-by download). The analytic result presented in \cite{XuTAAS2012} gives a sufficient condition (i.e., a specific parameter regime) under which the dynamics converge to a unique equilibrium, namely $\Pr(global\_state(t\to \infty)=0)=1$, meaning that all compromises will eventually be cleaned up. However, the properties of the dynamics in parameter regimes other than the specific regime characterized in \cite{XuTAAS2012} are not known until  \cite{XuTNSE2018}, which proves that the dynamics are {\em globally stable} in the {\em entire} parameter universe (i.e., the dynamics always converges to a unique equilibrium). This result remains true if the model parameters are extended to be node-dependent (i.e., different nodes $v\in V$ exhibit different cybersecurity characteristics, such as different host-based intrusion prevention/detection capabilities), and/or edge-dependent (i.e., different edges $e\in E$ exhibit different cybersecurity characteristics, such as different network-based intrusion prevention/detection capabilities) \cite{XuTNSE2018}. Moreover, the convergence speed is proven \cite{XuTNSE2018} to be exponential, except for a very special parameter regime (within which the dynamics converge polynomially). Although there is no closed-expression for the unique equilibrium, upper and lower bounds of the equilibrium can be obtained \cite{XuTAAS2012,XuTNSE2018}.
Another important insight, which shows the value of theoretic studies, is that there is a practical statistical method that can be used to estimate the global cybersecurity state at equilibrium {\em without} knowing the model parameters, thanks to the global stability of the dynamics \cite{XuTAAS2012,XuTNSE2018}.

The investigations mentioned above make the {\em independence} assumption that cyber attacks are waged independently of each other, which may not be the case when attacks are coordinated \cite{XuCollaborateCom08}. This highlights the importance of weakening, if not eliminating, the independence assumption.
Initial results have been reported in \cite{XuInternetMath2012,XuQuantitativeSecurityHotSoS2014,XuInternetMath2015Dependence}. An important finding is that assuming away the due dependence can lead to results that are unnecessarily restrictive, if not incorrect.
Since the aforementioned dependence can be caused by multiple cyber attackers,
preventive and reactive cyber defense dynamics have been extended to investigate the effect of multiple cyber attackers \cite{XuTDSC2012}, which may even fight against each other.
This leads to an interesting insight: the defender can leverage one attacker, say Alice, to ``defeat'' another attacker, say Bob, when the defender can more effectively defend against Alice than Bob.

In summary, we have a pretty deep understanding of preventive and reactive cyber defense dynamics.
For example, the effectiveness of preventive and reactive cyber defenses is limited by a fundamental attack-defense asymmetry:
the attack consequence is automatically amplified by a network effect reflected by
the largest eigenvalue (in modulus) of the attack-defense structure $G$; in contrast, the defense effectiveness is not amplified by any network effect.
This attack-defense asymmetry highlights the importance of enforcing strict network access control policy (e.g., direct communication between computers is allowed only when missions demand it), which effectively reduce the largest eigenvalue.

\subsection{Adaptive Cyber Defense Dynamics}

Cyber defense is often adaptive because the defender needs to adapt to the evolution of cyber attacks.
Adaptive cyber defense dynamics have been investigated in \cite{XuTAAS2014,XuQuantitativeSecurityHotSoS2014} while considering arbitrary attack-defense structure $G=(V,E)$.
In \cite{XuTAAS2014}, both semi-adaptive defenses (i.e., the defender dynamically adjusts the defense, but not necessarily geared towards the evolution of cyber attacks) and fully-adaptive defenses (i.e., the defender dynamically adjusts the defense geared towards the observed evolution of cyber attacks) are investigated.
Adaptive control strategies can be used to force the dynamics to follow a trajectory that benefits the defender (e.g., forcing the dynamics to converge to a certain equilibrium).
In \cite{XuQuantitativeSecurityHotSoS2014}, a new approach is proposed to model adaptive cyber defense dynamics with adaptive cyber attacks.
An interesting finding is that the global cybersecurity state is relatively easy to quantify when the defense is either highly effectively or highly ineffective.

In summary, both cyber attacks and defenses are often adaptive, but they are challenging to to model and analyze mathematically.
For example, the intuitive concept of {\em adaptation agility} needs to be systematically investigated, with an initial effort presented in \cite{XuAgiliy2018manuscript}.

\subsection{Proactive Cyber Defense Dynamics}

Adaptive defenses may rely on the successful detection of attacks. Proactive defense does not suffer from this restriction because the defender can adjust the defense regardless of whether there are successful attacks or not. Moving-Target Defense (MTD) is a popular example of proactive defense.
Many MTD techniques have been proposed (see, e.g., \cite{MIT-MTD-Survey-2013} and the numerous references therein) and many aspects of MTD have been investigated (see, e.g., \cite{Jafarian:2012:ORH:2342441.2342467,DBLP:conf/ccs/RahmanAB14,Maleki:2016:MMM:2995272.2995273,Connell:2017:PMM:3140549.3140550,10.1007/978-3-319-61176-1_2}). However, very few studies have aimed at systematically quantifying the effectiveness of MTD.  In what follows we outline a systematic use of MTD.

\begin{figure}[!htbp]
\centering
\includegraphics[width=.4\textwidth]{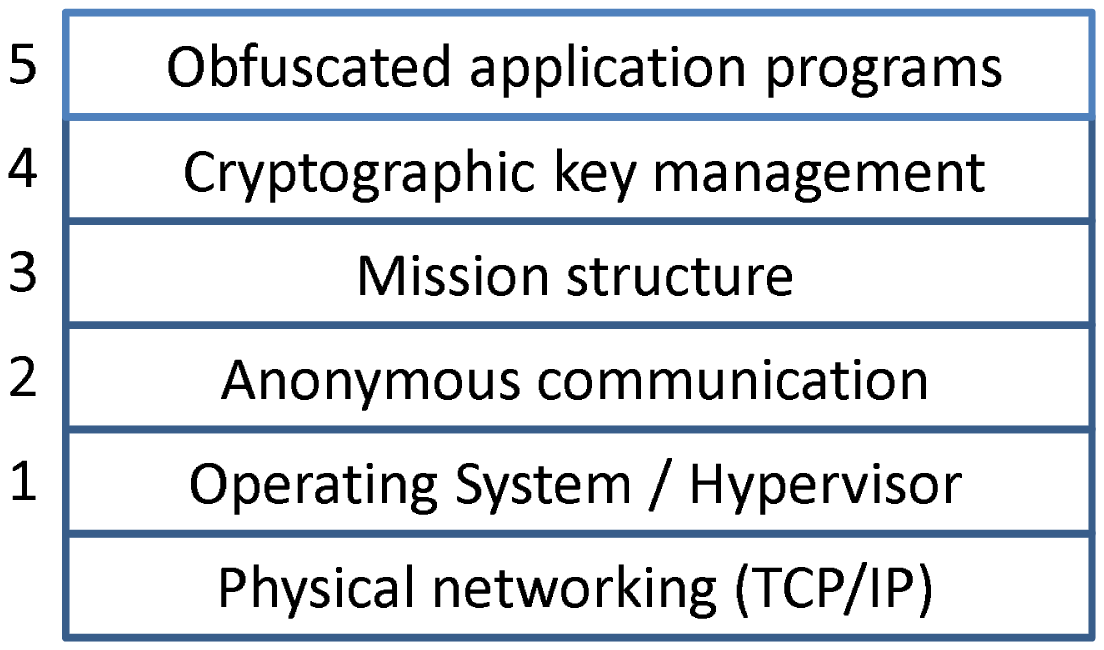}
\caption{An example architecture showing that MTD can be employed at different layers, individually or collectively.}
\label{fig:architecture}
\end{figure}

As highlighted in Figure \ref{fig:architecture}, MTD can be employed at one or multiple layers of the software stack.
Specifically, MTD can be employed at the operating system/hypervisor layer by frequently changing the underlying operating system/hypervisor environment (e.g., using VM migration).
Anonymous communication can be leveraged to disrupt the attacker's reconnaissance capabilities by degrading the attacker's capability from waging {\em targeted and adaptive attacks} to {\em random attacks} \cite{XuAsiaCCS08,XuIEEETIFS2011}. This means that anonymous communication can be leveraged for MTD to substantially increase the attacker's reconnaissance effort by dynamically adjusting, for example, the underlying anonymous communication infrastructure.
At the {\em mission structure} layer, ``mission structure'' may be represented by
a sub-graph $G_M(t)=(V_M(t),E_M(t))$ of the aforementioned attack-defense structure $G(t)=(V(t),E(t))$ with $V_M(t)\subseteq V(t)$ and $E_M(t)\subseteq E(t)$.
In order to prevent the attacker from identifying a target node (e.g., the cyber command-and-control center), the defender can frequently relocate the target node.
At the cryptographic key management layer, proactive cryptosystems \cite{HJJKY97}, key-insulated cryptosystems \cite{DKXY02,DKXY03,XuKIAsiaCCS2012}, or leak-free cryptosystems \cite{DingICDCS04,XuJCS09} can be used to tolerate the compromise of some computers, which hold some short-lived cryptographic key or cryptographic key shares \cite{DF89,XuAsiaCCS08,XuIEEETIFS2011}. Moreover, dynamic re-keying (e.g., \cite{DBLP:journals/jcs/Xu07,XUJCS06}) can be frequently enforced even in the absence of {\em detected} compromises because this can make the compromised cryptographic keys useless or can increase the chance that the compromise is detected \cite{XuFC09}. 
At the application layer, the defender can use the following kinds of MTD to slow down the attacker:
(i) re-obfuscating the application programs frequently; (ii) dynamically re-shuffle honeypot IP addresses within a production network to capture new attacks \cite{XuBookChapter13}.

While it is intuitive that MTD can be employed at each of these five layers, the main question is: When should the defender employ MTD and at which layers?
Towards answering this question, the first systematic quantification study is presented in \cite{XuHotSOS14-MTD}, which uses cybersecurity dynamics to quantify the effectiveness of MTD.
However, the investigation treats MTD as a means, rather than a goal. That is, the effectiveness of MTD is {\em indirectly}, rather than directly, measured in  \cite{XuHotSOS14-MTD}.
In summary, proactive cyber defense is one of the very few approaches that can potentially defend against sophisticated attacks, such as zero-day attacks and Advanced Persistent Threats (APTs).
More research needs to be done in order to systematically and directly quantify the effectiveness of proactive defense, including MTD.

\subsection{Active Cyber Defense Dynamics}

In the context of this chapter, active cyber defense means the use of ``defenseware'' (e.g., white worms or ``malware killer" programs) to detect and clean up compromised computers.
That is, active cyber defense is not about hacking back because it is employed within the administrative boundary of the network in question.
The systematic modeling study of active cyber defense dynamics is initiated in \cite{XuInternetMath2015ACD}, which formulates a mathematical model to quantify the effectiveness of active cyber defense. In active cyber defense dynamics, we need to consider a pair of attack-defense structures, denoted by $G_A(t)=(V_A(t),E_A(t))$ and $G_D(t)=(V_D(t),E_D(t))$. Note that $G_A(t)$ is centered at the attacker's point of view, and $G_D(t)$ is centered at the defender's point of view, while noting that it is possible that $G_A(t)=G_D(t)$.
This leads to the identification of the optimal $G_D(t)$ under certain circumstances.
In particular, it is shown \cite{XuInternetMath2015ACD} that active cyber defense can benefit the defender substantially by eliminating the aforementioned asymmetry, which is inherent to preventive and reactive cyber defense dynamics.

In \cite{XuGameSec13}, further investigation is conducted to identify optimal strategies for orchestrating active cyber defense against non-strategic or strategic attackers.
In order to effectively defend against a non-strategic attacker, two flavors of optimal control strategies are investigated (i.e., {\em infinite-time horizon} control vs. {\em fast} control),
by showing when the defender should adjust its active defense (including the extreme case of giving up the use of active defense, and instead using other kinds of defenses).
In order to effectively defend against a strategic attacker, we identify Nash equilibrium strategies, while considering factors such as whether or not the attacker is willing to expose its advanced or zero-day attacks (exposure implying likelihood that these attacks will soon become useless).

In \cite{XuHotSoS2015}, it is shown for the first time that active cyber defense dynamics can exhibit {\em bifurcation} and {\em chaos}. Their cybersecurity implications include
(i) it is not feasible or possible to seek to predict active cyber defense dynamics under certain circumstances, such as those reported in \cite{XuHotSoS2015};
(ii) the defender should seek to manipulate active cyber defense dynamics to avoid such ``unmanageable'' situations.
In summary, the defender can use active cyber defense to offset the asymmetry advantage of the attacker in preventive and reactive cyber defense dynamics.
However, active cyber defense is no panacea, and should be used together with other kinds of defenses \cite{XuGameSec13}.
Additional research needs to be conducted to deepen our understanding of active cyber defense dynamics.

\section{Cybersecurity Data Analytics}
\label{sec:data-analytics}

Like cybersecurity first-principle modeling and analysis, cybersecurity data analytics is also centered at some well-defined cybersecurity metrics. However, cybersecurity data analytics is complementary to the cybersecurity first-principle modeling and analysis because the former is data- and experiment-driven (rather than assumption- or semantics-driven). More specifically, cybersecurity data analytics aim to achieve a range of objectives, including: (i) obtaining model parameters used by cybersecurity first-principle models, (ii) validating or invalidating the assumptions made by cybersecurity first-principle models, and (iii) helping tackle the {\em transient behavior} of cybersecurity dynamics.
The state-of-the-art is that significant progress has been made in the aforementioned objectives (i) and (iii), which are reviewed below, but not in (ii) due to the lack of real-world datasets.

\subsection{Obtaining Model Parameters}

\noindent{\bf Measuring the attack-defense structure $G(t)$}.
In cybersecurity first-principle modeling and analysis, obtaining the attack-defense structure $G(t)$ is typically, and legitimately, treated as an orthogonal effort because it copes with a different aspect of the cybersecurity problem. As discussed above, researchers have recently started to investigate how to represent networks and computers at finer-grained granularities \cite{XuHotSoS2018Firewall,XuHotSoS2018Diversity}. Once a modeling resolution is determined, we need to represent the software stack (including applications and operating systems) of individual computers, represent individual computers as well as the dependence and communication relations within individual computers, represent the communication relations between computers (e.g., which computer or application is authorized to communicate with which other computer or application in a network), and represent the communication relation between a network and its external environment networks.

\smallskip

\noindent{\bf Measuring susceptibility of software systems}.
Cybersecurity first-principle models often assume parameters describing the {\em susceptibility} of an ``atom'' (e.g., computer or software component). In order to measure this parameter or metric, we need to measure the vulnerability of the ``atom''. From the perspective of software vulnerability, we need to measure to what extent a software program is vulnerable and susceptible to exploits. For this purpose, we need to understand and characterize the capabilities of {\em vulnerability detection} capabilities.
For example, static analysis of software source code is one approach to detecting vulnerabilities.
This approach can be further divided into two methods:
{\em code similarity-based} \cite{DBLP:conf/sp/KimWLO17,li2016vulpecker} vs. {\em pattern-based} \cite{FlawFinder,RATS,Checkmarx,grieco2016toward,neuhaus2007predicting,yamaguchi2013chucky,yamaguchi2012generalized,vuldeepecker,XuSySeVR2018}. The former method is effective in detecting vulnerabilities caused by certain kinds of code cloning \cite{vuldeepecker}. Pattern-based methods are not limited to detecting clone-caused vulnerabilities. Pattern-based detection methods detect vulnerabilities at a coarse granularity, such as at the level of individual programs \cite{grieco2016toward}, individual components \cite{neuhaus2007predicting}, individual files \cite{shin2011evaluating}, or individual functions \cite{yamaguchi2011vulnerability,yamaguchi2012generalized}.
More recent studies focus on fine-grained vulnerability detection \cite{DBLP:conf/sp/KimWLO17,li2016vulpecker,vuldeepecker,XuSySeVR2018}.
Studies in vulnerability detection represent a first step towards quantifying the susceptibility of software systems.

\smallskip

\noindent{\bf Measuring defense capabilities}.
The accurate measurement of defense capabilities is one outstanding open problem. For example, most existing measurements often assume the availability of the ground truth in question.
In the real world, ground truth is difficult to obtain. Therefore, it is important to investigate to what extent we can get rid of the ground truth, if possible at all.
In the context of evaluating the detection capabilities of malware detectors, this problem has been investigated in \cite{KantchelianACMAIS2016,XuTIFSTrustworthiness2018,XuMilcom2018sub-relativeaccuracy}. In particular, statistical estimators are designed and evaluated in \cite{XuTIFSTrustworthiness2018}. Moreover, relative accuracy of malware detectors, rather than absolute accuracy, can be estimated under much weaker assumptions \cite{XuMilcom2018sub-relativeaccuracy}.

\smallskip

\noindent{\bf Quantifying attack capabilities.}
In order to measure the capabilities of pull-based attacks (e.g., drive-by download), it is necessary to measure the extent at which malicious websites can evade detection systems.
There have been many proposals for detecting malicious websites (see, e.g., \cite{DBLP:journals/tist/MaSSV11,XuCODASPY13}). However, the open problem is that the attacker, who knows the detection model or the dataset from which the model is learned, can manipulate the malicious websites to evade the detection systems in question. The investigation of this problem is initiated in \cite{XuCNS2014}, but there are no satisfactory solutions yet. For example, the proactive training approach used in \cite{XuCNS2014} can only make the detection accuracy around 70-80\%, which is far from sufficient.

\subsection{Tackling the Transient Behavior Barrier}

\begin{figure}[!hbtp]
\centering
\includegraphics[width=.96\textwidth]{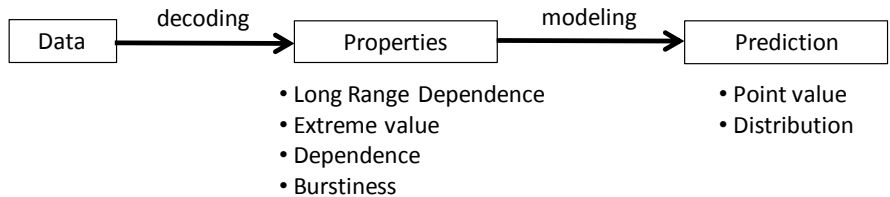}
\caption{The ``grey-box'' statistical methodology for cybersecurity data analytics.}
\label{fig:framework}
\end{figure}

Towards ultimately tackling the transient behavior barrier, Figure \ref{fig:framework} highlights the ``grey-box'' statistical methodology initiated in \cite{XuTIFS2013}. The term ``grey-box'' means that the methodology first aims to characterize the statistical properties exhibited by the data (e.g., long-range dependence, extreme value, dependence, burstiness), and then uses these properties to guide the development of prediction models.

\noindent{\bf Progress in coping with cybersecurity univariate time series}.
A particular kind of univariate time series, dubbed {\em stochastic cyber attack processes}, has been substantially investigated \cite{XuTIFS2013,XuTIFS2015,XuPLoSOne2015,XuMarkerPointProcess2017,XuTIFSDataBreach2018}. These cyber attack processes describe the number of cyber attacks or incidents against a target of interest (e.g., a network, a computer, or even a particular port).
Specifically, leveraging the statistical property known as {\em long-range dependence}, which is exhibited by the stochastic cyber attack processes corresponding to a dataset
collected at a honeypot, the ``grey-box'' methodology leads to an 80\% accuracy in forecasting the number of attacks coming to a network one hour ahead of time.
By further extending the model to accommodate the {\em extreme values} exhibited by the dataset, the one-hour ahead forecasting accuracy is improved to 88\% \cite{XuTIFS2015}.
A preliminary analysis of the spatiotemporal predictability shows that the forecasting upper bound is around 93\% \cite{XuPLoSOne2015}.
Focusing on the extreme values only, a {\em marked point process} model is developed to forecast the distribution of extreme values with good accuracy \cite{XuMarkerPointProcess2017}.
Another study focuses on the statistical analysis of breach incidents occurring between 2005-2017 \cite{XuTIFSDataBreach2018}, which shows that in contrast to previous beliefs,
both the inter-arrival times and the breach sizes of hacking breach incidents should be described using stochastic processes, rather than probabilistic distributions, because of the autocorrelations exhibited by the data. These properties can be exploited to build accurate forecasting models \cite{XuTIFSDataBreach2018}.
These results evidently show predictability in cyberspace, at least from the perspectives that have been explored.

\smallskip

\noindent{\bf Progress in coping with cybersecurity multivariate time series}.
Many cybersecurity datasets can be represented by multivariate time series. The first investigation of this kind is to characterize and forecast the effectiveness of cyber defense early-warnings
\cite{XuVineCopula2015}. The idea of early-warning is to filter the cyber attacks, which are detected at cyber defense instruments (e.g., honeypot \cite{ProvosUsenixSecurity04} or network telescope \cite{BaileyNDSS05}) or third parties \cite{XuBookChapter13}, against a network of interest. A unique research challenge, when compared with univariate time series, is to cope with the dependence between the time series, which manifests the dependence barrier \cite{XuCybersecurityDynamicsHotSoS2014} from a statistical perspective. For this purpose, the copula technique \cite{joe2014dependence} turns out to be useful.  A more general investigation of multivariate time series of cyber risks is conducted in \cite{XuJournalAppliedStatistics2018}.
The idea is to use a Copula-GARCH model to describe the multivariate dependence between stochastic cyber attack processes. In \cite{XuVineCopula2015,XuJournalAppliedStatistics2018}, it is shown that assuming away the due dependence between stochastic cyber attack processes (i.e., the time series) can cause a severe underestimation of cybersecurity risks.

\smallskip

\noindent{\bf Progress in coping with cybersecurity multivariate time series}.
Many cybersecurity datasets can be represented by graph time series. A concrete example is the reconnaissance behaviors of cyber attackers, which can be represented as a time series of bipartite graphs \cite{XuInTrust2014,XuBipartiteGraphTimeSeries2018}, which reflects one particular kind of the aforementioned attack-defense structure $G(t)=(V(t),E(t))$ over time $t$.
For studying such time series of graphs, a systematic methodology is presented in  \cite{XuBipartiteGraphTimeSeries2018}.
At a high level, the methodology is to characterize the evolution (i.e., time series) of the {\em similarity} between two adjacent graphs $G(t)$ and $G(t+1)$, where the notion of {\em similarity} can have many different definitions (leading to various kinds of analyses). Using a real-world dataset, it is shown, among other things, that a couple of time resolutions are sufficient to accommodate and describe the {\em temporal} characteristics of these time series. This finding offers an effective guideline in coping with real-time data streams of this kind in real-world defense operations.

\section{Future Research Directions}
\label{sec:future-research}

In this section we discuss future research directions with respect to the three axes mentioned above.

\subsection{Cybersecurity Metrics}

Towards ultimately tackling the problem cybersecurity metrics, the following two outstanding issues need to be resolved as soon as possible.
\begin{itemize}
\item Identifying a systematic, ideally complete, set of metrics that must be measured:
Although many metrics have been proposed in the literature \cite{Pendleton16}, the state-of-the-art is still that we do not know which metrics are essential to define and measure.
This is because most existing metrics are introduced simply because they can be measured; in contrast, we need to know what metrics must be measured \cite{Pfleeger:2009:UCM:1591880.1592054}. A fundamental question is: What kinds of metrics have to be measured in order to quantify cybersecurity?
Therefore, we need to know a systematic set of metrics that can adequately describe cybersecurity. Better yet, it is important to know if there is a {\em complete} set of metrics, where ``complete'' means that any metric of interest can be derived from this set of metrics. Moreover, it is important to investigate the cost for measuring each of these metrics. This is because if one metric is costly to measure, we may need to seek easy-to-measure, alternate metric(s) to replace the hard-to-measure one as long as the former can answer the same kinds of questions as the latter does.

\item Investigating mathematical properties of cybersecurity metrics and the operators that can be applied to them:
As mentioned above, cybersecurity can be characterized at multiple model resolutions, reminiscent of the idea of considering security at multiple layers of abstractions \cite{Lampson06practicalprinciples}. Ideally, cybersecurity metrics at a lower model resolution (i.e., a higher level of abstraction or macroscopic cybersecurity) should be a mathematical function of the cybersecurity metrics defined and measured at some higher model resolutions (i.e., lower levels of abstractions or microscopic cybersecurity). This incurs the issue of aggregating lower levels of cybersecurity metrics into higher levels metrics \cite{Pendleton16,5432146,Pfleeger:2009:UCM:1591880.1592054}. For this purpose, we need to investigate the mathematical properties that should be satisfied by cybersecurity metrics, including axiomatic properties.
\end{itemize}

\subsection{Cybersecurity First-Principle Modeling and Analysis}

The following unique set of technical barriers need to be adequately tackled.
\begin{itemize}
\item The {\em scalability} barrier \cite{XuCybersecurityDynamicsHotSoS2014}: A first-principle, native approach to modeling the evolution of global cybersecurity state caused by the attack-defense interactions would be Stochastic Process models, which would incur an exponentially-large state space that is not tractable in general. How can this problem be tackled while preserving information in the model as much as possible? The current approach is to use mean-field style treatment, which essentially reduces the number of dimensions from exponentially-many to a number of dimensions that is proportional to the size of the attack-defense structure $G(t)$.

\item The {\em nonlinearity} barrier \cite{XuCybersecurityDynamicsHotSoS2014}: It has been hypothesized that Cybersecurity Dynamics models are often highly nonlinear. The lack of real-world data has hindered the validation or rejection of this hypothesis, while many researchers believe in the nonlinearity. Coping with nonlinearity is a well known hard problem in general.

\item The {\em dependence} barrier \cite{XuCybersecurityDynamicsHotSoS2014}: The security states of the ``atoms'' are not independent of each other because, for example, some software may have the same vulnerabilities. It is an outstanding open problem to cope with the dependence between the security state of the ``atoms'' (i.e., random variables), for which initial progress has been made as mentioned above \cite{XuInternetMath2015Dependence,XuQuantitativeSecurityHotSoS2014}.

\item The {\em structural dynamics} barrier \cite{XuCybersecurityDynamicsHotSoS2014}: The attack-defense structure $G(t)$ itself evolves over time. There have been some studies on accommodating specific kind of evolutions (see, for example, \cite{XuHotSOS14-MTD}). However, we need to establish a mathematical description of general evolution of $G(t)$.

\item The {\em transient behavior} barrier \cite{XuCybersecurityDynamicsHotSoS2014}: Existing first-principle models often analyze the asymptotic behaviors of Cybersecurity Dynamics as time $t\to\infty$ (i.e., for sufficiently large $t$). For cybersecurity purposes, it is perhaps even more interesting to characterize the evolution of the global cybersecurity state before the dynamics converge to an equilibrium, if it does at all. This manifests the importance of cybersecurity data analytics.  Despite the progress reviewed above (e.g., \cite{XuTIFS2013,XuTIFS2015,XuPLoSOne2015,XuMarkerPointProcess2017,XuTIFSDataBreach2018}), our understanding of the problem is still at the infant stage.

\item The {\em uncertainty} barrier: Cybersecurity first-principle modeling often assumes the availability of complete information, meaning that the model parameters can be obtained precisely. This represents a first-step in building analytic models for baseline understanding. In practice, model parameters may not be known or may not be precisely measured, which highlights the importance of quantifying the consequences caused by uncertainties in the models and/or parameters.

\item The {\em deception} barrier: In the cybersecurity domain, data or information not only can be missing or noisy, but also can be malicious because the attacker can intentionally inject or manipulate the measurements in question to mislead the defenders. This kind of deceptive data/information needs to be rigorously treated.

\item The {\em human factor} barrier: The degrees that human users or defenders are vulnerable to social-engineering attacks need to be measured and quantified.
\end{itemize}

\subsection{Cybersecurity Data Analytics}

The following research problems need to be adequately resolved as soon as possible.
\begin{itemize}
\item Building a full-fledged statistical methodology to forecast holistic cybersecurity situational awareness,
including the the emergence of software zero-day vulnerabilities and attacks exploiting them.
Although the studies reviewed above already showed the feasibility of predicting cybersecurity situational awareness from certain specific perspectives
\cite{XuTIFS2013,XuInTrust2014,XuTIFS2015,XuPLoSOne2015,XuVineCopula2015,XuMarkerPointProcess2017,XuJournalAppliedStatistics2018,XuTIFSDataBreach2018},
these results only represent a first step towards the ultimate goal.

\item Tackling the dependence barrier as manifested in cybersecurity data analytics.
Cybersecurity data can have extremely high dimensions, while dependence can be inherent to them.
Therefore, we need to investigate forecasting models that can adequately accommodate the dependence ``encoded'' in real-world data.
The results mentioned above \cite{XuVineCopula2015,XuJournalAppliedStatistics2018} only address a small tip of the iceberg.
\end{itemize}

\section{Related Work}
\label{sec:related-work}

\noindent{\bf Prior studies related to the Cybersecurity Dynamics foundation}.
The present chapter systematically refines and extends an earlier treatment of Cybersecurity Dynamics \cite{XuCybersecurityDynamicsHotSoS2014}, while accommodating the many advancements during the past few years. Although there have been investigations on exploring the various aspects (or characteristics) of the science of cybersecurity
\cite{7958573,DBLP:conf/hotsos/RoqueBD16,Schneider11_BlueprintForScienceOfCybersecurity,DBLP:conf/nspw/SpringMP17,Kott2014}, to the best of our knowledge, we are the first to systematically map out a concrete framework as reviewed in the present chapter.

\smallskip

\noindent{\bf Prior studies related to cybersecurity metrics}.
There are several recent surveys related to cybersecurity metrics \cite{Pendleton16,8017389,Noel2017,XuSTRAM2018manuscript}.
Moreover, the problem has been rejuvenated by new efforts \cite{Pendleton16,8017389,Noel2017,Cho16-milcom,XuHotSoS2018Firewall,XuHotSoS2018Diversity,XuAgiliy2018manuscript,XuSTRAM2018manuscript}.
We treat cybersecurity metrics systematically, as highlighted in Eq. \eqref{eq:overall-equation}, for describing the configurations of networks, for describing systems and human vulnerabilities, for describing defense postures, for describing cyber attacks (i.e., threat models), and for describing the global cybersecurity state or cybersecurity situational awareness.

\smallskip

\noindent{\bf Prior studies related to cybersecurity first-principle modeling and analysis}.
As discussed in \cite{XuCybersecurityDynamicsHotSoS2014}, cybersecurity first-principle modeling is inspired by multiple endeavors in several disciplines, including:
(i) Biological epidemic models \cite{McKendrick1926,Kermack1927,Bailey1975,Anderson1991,HethcoteSIAMRew00}: These models have been adapted to the Internet setting
(or cyber epidemic models) since Kephart and White \cite{KephartOkland91,KephartOkland93}. Later efforts aim to accommodate general network structures,
including power-law network structures \cite{Pastor2001,Moreno2002,Pastor2001,Pastor2002a,NewmanReview2003,Vespignanibook08}
and {\em arbitrary} network structures (e.g., \cite{WangSRDS03,TowsleyInfocom05,WangTISSEC08,VanMieghemIEEEACMTON09}).
(ii) Interacting particle systems \cite{Liggett1985}: These models investigate the collective behaviors of interacting components and the phenomena that can emerge from these interactions.
(iii) Microfoundation in Economics \cite{Hoover2010}: This effort aims to make connections between macroeconomic models to the underlying microeconomic models.
However, the aforementioned technical barriers distinguish cybersecurity first-principle models from the models in the literature mentioned above.
Moreover, it is the Cybersecurity Dynamics foundation that stresses that the attack-defense structure reflects, among other things, the access control policies that are enforced in a network,
rather than the physical communication network. Furthermore, the foundation offers a unique set of cyber defense dynamics as reviewed above: preventive and reactive cyber defense dynamics,
adaptive cyber defense dynamics, proactive cyber defense dynamics, and active cyber defense dynamics.

It is worth mentioning that cybersecurity first-principle models in the context of Cybersecurity Dynamics are different from the models in the context of Attack Graphs (see, for example,
\cite{Phillips:1998:GSN:310889.310919,JhaOkland02,Ritchey:2000:UMC:882494.884423,Jha:2002:TFA:794201.795177,Ammann:2002:SGN:586110.586140,JajodiaDSN2012,Homer:2013:AVM:2590624.2590627,OuCyberSA2014}). This is because models in the context of Attack Graphs are {\em combinatorial} in nature (e.g., computing or enumerating attack paths with respect to a target); in contrast,
models in the context of Cybersecurity Dynamics are {\em stochastic processes} in nature because they explicitly model the evolution of the global cybersecurity state over time $t$.
This explains why these models are, as mentioned above, inspired by Biological Epidemic Models, Interacting Particle Systems, and Microfoundation in Economics, and why we can model various kinds of cyber defense dynamics.

\smallskip

\noindent{\bf Prior studies related to cybersecurity data analytics}.
There are numerous data-driven cybersecurity research activities, which however are often geared towards some specific events, attacks, or defenses.
For example, honeypot-captured cyber attack data have been used for purposes of visualization \cite{HerreroIJNS12},
clustering attacks \cite{new.att.honeypot,honeypot.pca,iat.cliques,vis.att}, and characterizing attack behaviors such as inter-arrival times \cite{DBLP:series/ais/AlataDDKKNPP06,DacierCoRR07}.
In contrast, cybersecurity data analytics in the context of the present chapter is meant to become an inherent pillar of the Cybersecurity Dynamics foundation, by interacting with the other two pillars (i.e., cybersecurity first-principle modeling and analysis and cybersecurity metrics) as shown in Figure \ref{fig:three-axis-relationship}.

\section{Conclusion}
\label{sec:conclusion}

We have systematically reviewed the Cybersecurity Dynamics foundation, with emphasis on the three active research axes or pillars in
cybersecurity metrics, cybersecurity first-principle modeling and analysis, and cybersecurity data analytics. We discussed the progress in each of these axes and future research directions.
We hope that we have clearly and successfully conveyed the following message: This is an exciting, but challenging, research endeavor that deserves a community effort to explore. We hope the present chapter will inspire many more studies towards achieving the ultimate, full-fledged Cybersecurity Dynamics foundation for the Science of Cybersecurity.

%\begin{acknowledgement}
\noindent{\bf Acknowledgement}.
This work was supported in part by ARO Grant \# W911NF-17-1-0566 and ARL Grant \# W911NF-17-2-0127.
The author would like to thank his mentors for their encouragement, and his collaborators (including his former and current students)
for deepening his understanding of the problem and potential solutions.
The author would also like to thank Lisa Ho and John Charlton for proofreading the present chapter.
%\end{acknowledgement}

%\bibliographystyle{plain}
%\bibliography{d:/my-data/papers/metrics,d:/my-data/papers/metrics-bib-new,d:/my-data/papers/metrics-richard,d:/my-data/papers/honeypot-new,d:/my-data/papers/honeypot,d:/my-data/papers/vulnerabilities,d:/my-data/papers/trustworthiness,d:/my-data/papers/measurement,d:/my-data/papers/mtd,d:/my-data/papers/botnet,d:/my-data/papers/crypto,d:/my-data/papers/complex-network,d:/my-data/papers/reference,d:/my-data/papers/percolation}

\begin{thebibliography}{100}

\bibitem{RATS}
{Rough Audit Tool for Security}, 2014.
\newblock
  \url{https://code.google.com/archive/p/rough-auditing-tool-for-security/}.

\bibitem{Checkmarx}
Checkmarx, 2018.
\newblock \url{https://www.checkmarx.com/}.

\bibitem{FlawFinder}
Flawfinder, 2018.
\newblock \url{http://www.dwheeler.com/flawfinder}.

\bibitem{DBLP:series/ais/AlataDDKKNPP06}
E.~Alata, M.~Dacier, Y.~Deswarte, M.~Kaa{\^a}niche, K.~Kortchinsky,
  V.~Nicomette, V.~Pham, and F.~Pouget.
\newblock Collection and analysis of attack data based on honeypots deployed on
  the internet.
\newblock In {\em Proc. Quality of Protection - Security Measurements and
  Metrics}, pages 79--91, 2006.

\bibitem{JajodiaDSN2012}
Massimiliano Albanese, Sushil Jajodia, and Steven Noel.
\newblock Time-efficient and cost-effective network hardening using attack
  graphs.
\newblock In {\em Proc. IEEE DSN'12}, pages 1--12, 2012.

\bibitem{AlbertNature00}
R.~Albert, H.~Jeong, and A.~Barabasi.
\newblock Error and attack tolerance of complex networks.
\newblock {\em Nature}, 406:378--482, 2000.

\bibitem{iat.cliques}
S.~Almotairi, A.~Clark, M.~Dacier, C.~Leita, G.~Mohay, V.~Pham, O.~Thonnard,
  and J.~Zimmermann.
\newblock Extracting inter-arrival time based behaviour from honeypot traffic
  using cliques.
\newblock In {\em 5th Australian Digital Forensics Conference}, pages 79--87,
  2007.

\bibitem{honeypot.pca}
S.~Almotairi, A.~Clark, G.~Mohay, and J.~Zimmermann.
\newblock Characterization of attackers' activities in honeypot traffic using
  principal component analysis.
\newblock In {\em Proc. IFIP International Conference on Network and Parallel
  Computing}, pages 147--154, 2008.

\bibitem{new.att.honeypot}
S.~Almotairi, A.~Clark, G.~Mohay, and J.~Zimmermann.
\newblock A technique for detecting new attacks in low-interaction honeypot
  traffic.
\newblock In {\em Proc. International Conference on Internet Monitoring and
  Protection}, pages 7--13, 2009.

\bibitem{Ammann:2002:SGN:586110.586140}
Paul Ammann, Duminda Wijesekera, and Saket Kaushik.
\newblock Scalable, graph-based network vulnerability analysis.
\newblock In {\em Proc. ACM CCS'02}, pages 217--224.

\bibitem{Anderson1991}
R.~Anderson and R.~May.
\newblock {\em Infectious Diseases of Humans}.
\newblock Oxford University Press, 1991.

\bibitem{BaileyNDSS05}
M.~Bailey, E.~Cooke, F.~Jahanian, J.~Nazario, and D.~Watson.
\newblock Internet motion sensor: A distributed blackhole monitoring system.
\newblock In {\em Proceedings of The 12th Network and Distributed System
  Security Symposium (NDSS'05)}, 2005.

\bibitem{Bailey1975}
N.~Bailey.
\newblock {\em The Mathematical Theory of Infectious Diseases and Its
  Applications}.
\newblock 2nd Edition. Griffin, London, 1975.

\bibitem{Vespignanibook08}
A.~Barrat, M.~Barthlemy, and A.~Vespignani.
\newblock {\em Dynamical Processes on Complex Networks}.
\newblock Cambridge University Press, 2008.

\bibitem{WangTISSEC08}
D.~Chakrabarti, Y.~Wang, C.~Wang, J.~Leskovec, and C.~Faloutsos.
\newblock Epidemic thresholds in real networks.
\newblock {\em ACM Trans. Inf. Syst. Secur.}, 10(4):1--26, 2008.

\bibitem{XuMilcom2018sub-relativeaccuracy}
John Charlton, Pang Du, Jin-Hee Cho, and Shouhuai Xu.
\newblock Measuring relative accuracy of malware detectors in the absence of
  ground truth.
\newblock manuscript under review, 2018.

\bibitem{XuESORICS2018sub}
Haoyu Chen, Deqing Zou, Shouhuai Xu, Hai Jin, Bin Yuan, and Yu~Lu.
\newblock Audy: Towards automated generation and deployment of dynamic network
  security rules.
\newblock manuscript under review, 2018.

\bibitem{XuHotSoS2018Firewall}
Huashan Chen, Jin{-}Hee Cho, and Shouhuai Xu.
\newblock Quantifying the security effectiveness of firewalls and dmzs.
\newblock In {\em Proceedings of the 5th Annual Symposium and Bootcamp on Hot
  Topics in the Science of Security (HoTSoS'2018)}, pages 9:1--9:11, 2018.

\bibitem{XuHotSoS2018Diversity}
Huashan Chen, Jin{-}Hee Cho, and Shouhuai Xu.
\newblock Quantifying the security effectiveness of network diversity: poster.
\newblock In {\em Proceedings of the 5th Annual Symposium and Bootcamp on Hot
  Topics in the Science of Security (HoTSoS'2018)}, page 24:1, 2018.

\bibitem{XuPLoSOne2015}
Yu-Zhong Chen, Zi-Gang Huang, Shouhuai Xu, and Ying-Cheng Lai.
\newblock Spatiotemporal patterns and predictability of cyberattacks.
\newblock {\em PLoS One}, 10(5):e0124472, 05 2015.

\bibitem{OuCyberSA2014}
Yi~Cheng, Julia Deng, Jason Li, ScottA. DeLoach, Anoop Singhal, and Xinming Ou.
\newblock Metrics of security.
\newblock In {\em Cyber Defense and Situational Awareness}, volume~62. 2014.

\bibitem{XuSTRAM2018manuscript}
J.~Cho, S.~Xu, P.~Hurley, M.~Mackay, T.~Benjamin, and M.~Beaumont.
\newblock Stram: Measuring the trustworthiness of computer-based systems.
\newblock manuscript in submission, 2018.

\bibitem{Cho16-milcom}
Jin-Hee Cho, Packtrick Hurley, and Shouhuai Xu.
\newblock Metrics and measurement of trustworthy systems.
\newblock In {\em IEEE Military Communication Conference (MILCOM 2016)}, 2016.

\bibitem{ChowUsenixSecurity04}
J.~Chow, B.~Pfaff, T.~Garfinkel, K.~Christopher, and M.~Rosenblum.
\newblock Understanding data lifetime via whole system simulation.
\newblock In {\em Proceedings of Usenix Security Symposium 2004}, 2004.

\bibitem{Connell:2017:PMM:3140549.3140550}
Warren Connell, Daniel~A. Menasc{\'e}, and Massimiliano Albanese.
\newblock Performance modeling of moving target defenses.
\newblock In {\em Proceedings of the 2017 Workshop on Moving Target Defense},
  MTD '17, pages 53--63, 2017.

\bibitem{vis.att}
G.~Conti and K.~Abdullah.
\newblock Passive visual fingerprinting of network attack tools.
\newblock In {\em Proc. 2004 ACM workshop on Visualization and data mining for
  computer security}, pages 45--54, 2004.

\bibitem{IRC-hardproblemlist}
INFOSEC~Research Council.
\newblock Hard problem list.
\newblock
  \url{http://www.infosec-research.org/docs_public/20051130-IRC-HPL-FINAL.pdf},
  2007.

\bibitem{XuQuantitativeSecurityHotSoS2014}
Gaofeng Da, Maochao Xu, and Shouhuai Xu.
\newblock A new approach to modeling and analyzing security of networked
  systems.
\newblock In {\em Proceedings of the 2014 Symposium and Bootcamp on the Science
  of Security (HotSoS'14)}, pages 6:1--6:12, 2014.

\bibitem{DBLP:conf/acsac/DagonGLL07}
David Dagon, Guofei Gu, Christopher~P. Lee, and Wenke Lee.
\newblock A taxonomy of botnet structures.
\newblock In {\em 23rd Annual Computer Security Applications Conference
  (ACSAC'07)}, pages 325--339, 2007.

\bibitem{DF89}
Y.~Desmedt and Y.~Frankel.
\newblock Threshold cryptosystems.
\newblock In {\em Proc.\ CRYPTO 89}, pages 307--315, 1989.

\bibitem{DingICDCS04}
X.~Ding, G.~Tsudik, and S.~Xu.
\newblock Leak-free group signatures with immediate revocation.
\newblock In {\em 24th International Conference on Distributed Computing
  Systems (ICDCS 2004)}, pages 608--615. IEEE Computer Society, 2004.

\bibitem{XuJCS09}
Xuhua Ding, Gene Tsudik, and Shouhuai Xu.
\newblock Leak-free mediated group signatures.
\newblock {\em Journal of Computer Security}, 17(4):489--514, 2009.

\bibitem{DKXY02}
Y.~Dodis, J.~Katz, S.~Xu, and M.~Yung.
\newblock Key-insulated public key cryptosystems.
\newblock In Lars~R. Knudsen, editor, {\em Advances in Cryptology - EUROCRYPT
  2002}, volume 2332 of {\em Lecture Notes in Computer Science}, pages 65--82.
  Springer, 2002.

\bibitem{DKXY03}
Y.~Dodis, J.~Katz, S.~Xu, and M.~Yung.
\newblock Strong key-insulated signature schemes.
\newblock In {\em Public Key Cryptography (PKC'03)}, pages 130--144, 2003.

\bibitem{XuKIAsiaCCS2012}
Yevgeniy Dodis, Weiliang Luo, Shouhuai Xu, and Moti Yung.
\newblock Key-insulated symmetric key cryptography and mitigating attacks
  against cryptographic cloud software.
\newblock In {\em 7th {ACM} Symposium on Information, Compuer and
  Communications Security, {ASIACCS} '12}, pages 57--58, 2012.

\bibitem{XuTIFSTrustworthiness2018}
P.~Du, Z.~Sun, H.~Chen, J.~H. Cho, and S.~Xu.
\newblock Statistical estimation of malware detection metrics in the absence of
  ground truth.
\newblock {\em IEEE Transactions on Information Forensics and Security}, pages
  1--1, 2018.

\bibitem{TowsleyInfocom05}
A.~Ganesh, L.~Massoulie, and D.~Towsley.
\newblock The effect of network topology on the spread of epidemics.
\newblock In {\em Proceedings of IEEE Infocom 2005}, 2005.

\bibitem{XuComplexNetworkSub2018}
Richard Garcia-Lebron, David~J. Myers, Shouhuai Xu, and Jie Sun.
\newblock Node diversification in complex networks by decentralized coloring).
\newblock manuscript under review, 2018.

\bibitem{XuBipartiteGraphTimeSeries2018}
Richard Garcia-Lebron, Kristin Schweitzer, Raymond Bateman, and Shouhuai Xu.
\newblock A framework for characterizing the evolution of cyber attacker-victim
  relation graphs.
\newblock manuscript in submission, 2018.

\bibitem{grieco2016toward}
Gustavo Grieco, Guillermo~Luis Grinblat, Lucas~C. Uzal, Sanjay Rawat, Josselin
  Feist, and Laurent Mounier.
\newblock Toward large-scale vulnerability discovery using machine learning.
\newblock In {\em Proceedings of the Sixth {ACM} on Conference on Data and
  Application Security and Privacy, {CODASPY} 2016, New Orleans, LA, USA},
  pages 85--96, 2016.

\bibitem{Guan:2015:PPK:2867539.2867702}
Le~Guan, Jingqiang Lin, Bo~Luo, Jiwu Jing, and Jing Wang.
\newblock Protecting private keys against memory disclosure attacks using
  hardware transactional memory.
\newblock In {\em Proceedings of the 2015 IEEE Symposium on Security and
  Privacy}, SP '15, pages 3--19, 2015.

\bibitem{XuHotSOS14-MTD}
Yujuan Han, Wnelian Lu, and Shouhuai Xu.
\newblock Characterizing the power of moving target defense via cyber epidemic
  dynamics.
\newblock In {\em Proc. 2014 Symposium and Bootcamp on the Science of Security
  (HotSoS'14)}, pages 10:1--10:12, 2014.

\bibitem{HarrisonDSN07}
K.~Harrison and S.~Xu.
\newblock Protecting cryptographic keys from memory disclosures.
\newblock In {\em Proceedings of the 2007 IEEE/IFIP International Conference on
  Dependable Systems and Networks (DSN-DCCS'07)}, pages 137--143. IEEE Computer
  Society, 2007.

\bibitem{7958573}
C.~Herley and P.~C. v.~Oorschot.
\newblock Sok: Science, security and the elusive goal of security as a
  scientific pursuit.
\newblock In {\em 2017 IEEE Symposium on Security and Privacy (SP)}, pages
  99--120, May 2017.

\bibitem{HerreroIJNS12}
A.~Herrero, U.~Zurutuza, and E.~Corchado.
\newblock A neural-visualization ids for honeynet data.
\newblock {\em Int. J. Neural Syst.}, 22(2), 2012.

\bibitem{HJJKY97}
A.~Herzberg, M.~Jakobsson, S.~Jarecki, H.~Krawczyk, and M.~Yung.
\newblock Proactive public key and signature schemes.
\newblock In {\em Proceedings of the Fourth Annual Conference on Computer and
  Communications Security}, pages 100--110. ACM, 1997.

\bibitem{HethcoteSIAMRew00}
H.~Hethcote.
\newblock The mathematics of infectious diseases.
\newblock {\em SIAM Rev.}, 42(4):599--653, 2000.

\bibitem{Homer:2013:AVM:2590624.2590627}
John Homer, Su~Zhang, Xinming Ou, David Schmidt, Yanhui Du, S.~Raj Rajagopalan,
  and Anoop Singhal.
\newblock Aggregating vulnerability metrics in enterprise networks using attack
  graphs.
\newblock {\em J. Comput. Secur.}, 21(4):561--597, 2013.

\bibitem{Hoover2010}
K.~Hoover.
\newblock Idealizing reduction: The microfoundations of macroeconomics.
\newblock {\em Erkenntnis}, 73:329--347, 2010.

\bibitem{HussainSIGCOMM03}
A.~Hussain, J.~Heidemann, and C.~Papadopoulos.
\newblock A framework for classifying denial of service attacks.
\newblock In {\em Proceedings of ACM SIGCOMM'03}, pages 99--110, 2003.

\bibitem{CyberKillChainPaper2011}
Eric~M. Hutchins, Michael~J. Cloppert, and Rohan~M. Amin.
\newblock Intelligence-driven computer network defense informed by analysis of
  adversary campaigns and intrusion kill chains.
\newblock In {\em 2011 International Conference on Information Warfare and
  Security}.

\bibitem{Jafarian:2012:ORH:2342441.2342467}
Jafar~Haadi Jafarian, Ehab Al-Shaer, and Qi~Duan.
\newblock Openflow random host mutation: Transparent moving target defense
  using software defined networking.
\newblock In {\em Proceedings of the First Workshop on Hot Topics in Software
  Defined Networks (HotSDN'12)}, pages 127--132, 2012.

\bibitem{Jha:2002:TFA:794201.795177}
S.~Jha, O.~Sheyner, and J.~Wing.
\newblock Two formal analys s of attack graphs.
\newblock In {\em Proc. IEEE Workshop on Computer Security Foundations}, pages
  49--59, 2002.

\bibitem{joe2014dependence}
Harry Joe.
\newblock {\em Dependence Modeling with Copulas}.
\newblock CRC Press, 2014.

\bibitem{DBLP:conf/ccs/JuelsK07}
Ari Juels and Burton S.~Kaliski Jr.
\newblock Pors: proofs of retrievability for large files.
\newblock In {\em Proc. ACM Conference on Computer and Communications Security
  (CCS'07)}, pages 584--597, 2007.

\bibitem{DacierCoRR07}
M.~Ka{\^a}niche, Y.~Deswarte, E.~Alata, M.~Dacier, and V.~Nicomette.
\newblock Empirical analysis and statistical modeling of attack processes based
  on honeypots.
\newblock {\em CoRR}, abs/0704.0861, 2007.

\bibitem{KantchelianACMAIS2016}
Alex Kantchelian, Michael~Carl Tschantz, Sadia Afroz, Brad Miller, Vaishaal
  Shankar, Rekha Bachwani, Anthony~D Joseph, and J~Doug Tygar.
\newblock Better malware ground truth: Techniques for weighting anti-virus
  vendor labels.
\newblock In {\em Proceedings of the 8th ACM Workshop on Artificial
  Intelligence and Security}, pages 45--56. ACM, 2015.

\bibitem{XuACNS2010}
Erhan~J. Kartaltepe, Jose~Andre Morales, Shouhuai Xu, and Ravi~S. Sandhu.
\newblock Social network-based botnet command-and-control: Emerging threats and
  countermeasures.
\newblock In {\em ACNS}, pages 511--528, 2010.

\bibitem{KephartOkland91}
J.~Kephart and S.~White.
\newblock Directed-graph epidemiological models of computer viruses.
\newblock In {\em IEEE Symposium on Security and Privacy}, pages 343--361,
  1991.

\bibitem{KephartOkland93}
J.~Kephart and S.~White.
\newblock Measuring and modeling computer virus prevalence.
\newblock In {\em IEEE Symposium on Security and Privacy}, pages 2--15, 1993.

\bibitem{Kermack1927}
W.~Kermack and A.~McKendrick.
\newblock A contribution to the mathematical theory of epidemics.
\newblock {\em Proc. of Roy. Soc. Lond. A}, 115:700--721, 1927.

\bibitem{DBLP:conf/sp/KimWLO17}
Seulbae Kim, Seunghoon Woo, Heejo Lee, and Hakjoo Oh.
\newblock {VUDDY:} {A} scalable approach for vulnerable code clone discovery.
\newblock In {\em 2017 {IEEE} Symposium on Security and Privacy}, pages
  595--614, 2017.

\bibitem{DBLP:journals/corr/abs-1801-01203}
Paul Kocher, Daniel Genkin, Daniel Gruss, Werner Haas, Mike Hamburg, Moritz
  Lipp, Stefan Mangard, Thomas Prescher, Michael Schwarz, and Yuval Yarom.
\newblock Spectre attacks: Exploiting speculative execution.
\newblock {\em CoRR}, abs/1801.01203, 2018.

\bibitem{kopf2007information}
Boris K{\"o}pf and David Basin.
\newblock An information-theoretic model for adaptive side-channel attacks.
\newblock In {\em Proc. ACM conference on Computer and communications
  security}, pages 286--296. ACM, 2007.

\bibitem{Kott2014}
Alexander Kott.
\newblock Towards fundamental science of cyber security.
\newblock In Robinson~E. Pino, editor, {\em Network Science and Cybersecurity},
  volume~55 of {\em Advances in Information Security}, pages 1--13. Springer
  New York, 2014.

\bibitem{Lampson06practicalprinciples}
Butler Lampson.
\newblock Practical principles for computer security, 2006.

\bibitem{DBLP:conf/IEEEares/LeonardXS09a}
Justin Leonard, Shouhuai Xu, and Ravi~S. Sandhu.
\newblock A framework for understanding botnets.
\newblock In {\em Proceedings of the The Forth International Conference on
  Availability, Reliability and Security, {ARES} 2009}, pages 917--922, 2009.

\bibitem{XuAINA07}
X.~Li, P.~Parker, and S.~Xu.
\newblock Towards quantifying the (in)security of networked systems.
\newblock In {\em 21st IEEE International Conference on Advanced Information
  Networking and Applications (AINA'07)}, pages 420--427, 2007.

\bibitem{XuTDSC2011}
X.~Li, P.~Parker, and S.~Xu.
\newblock A stochastic model for quantitative security analysis of networked
  systems.
\newblock {\em IEEE Transactions on Dependable and Secure Computing},
  8(1):28--43, 2011.

\bibitem{vuldeepecker}
Z.~{Li}, D.~{Zou}, S.~{Xu}, X.~{Ou}, H.~{Jin}, S.~{Wang}, Z.~{Deng}, and
  Y.~{Zhong}.
\newblock Vuldeepecker: A deep learning-based system for vulnerability
  detection.
\newblock In {\em Proceedings of the 25th Annual Network and Distributed System
  Security Symposium (NDSS'2018)}, 2018.

\bibitem{li2016vulpecker}
Zhen Li, Deqing Zou, Shouhuai Xu, Hai Jin, Hanchao Qi, and Jie Hu.
\newblock Vulpecker: An automated vulnerability detection system based on code
  similarity analysis.
\newblock In {\em Proceedings of the 32nd Annual Conference on Computer
  Security Applications, {ACSAC} 2016, Los Angeles, CA, USA}, pages 201--213,
  2016.

\bibitem{XuSySeVR2018}
Zhen Li, Deqing Zou, Shouhuai Xu, Hai Jin, Yawei Zhu, Zhaoxuan Chen, Sujuan
  Wang, and Jialai Wang.
\newblock Sysevr: A framework for using deep learning to detect software
  vulnerabilities.
\newblock Manuscript in submission, 2018.

\bibitem{Liggett1985}
T.~Liggett.
\newblock {\em Interacting Particle Systems}.
\newblock Springer, 1985.

\bibitem{XuGameSec13}
Wenlian Lu, Shouhuai Xu, and Xinlei Yi.
\newblock Optimizing active cyber defense dynamics.
\newblock In {\em Proceedings of the 4th International Conference on Decision
  and Game Theory for Security (GameSec'13)}, pages 206--225, 2013.

\bibitem{XuBookChapter13}
Weiliang Luo, Li~Xu, Zhenxin Zhan, Qingji Zheng, and Shouhuai Xu.
\newblock Federated cloud security architecture for secure and agile clouds.
\newblock In Keesook~J. Han, Baek-Young Choi, and Sejun Song, editors, {\em
  High Performance Cloud Auditing and Applications}, pages 169--188. Springer
  New York, 2014.

\bibitem{DBLP:journals/tist/MaSSV11}
Justin Ma, Lawrence~K. Saul, Stefan Savage, and Geoffrey~M. Voelker.
\newblock Learning to detect malicious urls.
\newblock {\em {ACM} {TIST}}, 2(3):30:1--30:24, 2011.

\bibitem{Maleki:2016:MMM:2995272.2995273}
Hoda Maleki, Saeed Valizadeh, William Koch, Azer Bestavros, and Marten van
  Dijk.
\newblock Markov modeling of moving target defense games.
\newblock In {\em Proceedings of the 2016 ACM Workshop on Moving Target
  Defense}, MTD '16, pages 81--92, 2016.

\bibitem{Mandiant}
Mandiant.
\newblock Apt1 report.
\newblock
  \url{https://www.fireeye.com/content/dam/fireeyewww/services/pdfs/mandiant-apt1-report.pdf},
  February 16, 2013 (Accessed July 08, 2016).

\bibitem{McKendrick1926}
A.~McKendrick.
\newblock Applications of mathematics to medical problems.
\newblock {\em Proc. of Edin. Math. Soceity}, 14:98--130, 1926.

\bibitem{XuAgiliy2018manuscript}
Jose Mireles, Eric Ficke, Jin-Hee Cho, Patrick Hurley, and Shouhuai Xu.
\newblock Metrics towards measuring cyber agility.
\newblock manuscript in submission, 2018.

\bibitem{DBLP:conf/dimva/MohaisenA14}
Aziz Mohaisen and Omar Alrawi.
\newblock Av-meter: An evaluation of antivirus scans and labels.
\newblock In {\em Detection of Intrusions and Malware, and Vulnerability
  Assessment - 11th International Conference, {DIMVA} 2014, Proceedings}, pages
  112--131, 2014.

\bibitem{XuASE2012}
Jose Morales, Shouhuai Xu, and Ravi Sandhu.
\newblock Analyzing malware detection efficiency with multiple anti-malware
  programs.
\newblock In {\em Proceedings of 2012 ASE International Conference on Cyber
  Security (CyberSecurity'12)}, 2012.

\bibitem{Moreno2002}
Y.~Moreno, R.~Pastor-Satorras, and A.~Vespignani.
\newblock Epidemic outbreaks in complex heterogeneous networks.
\newblock {\em European Physical Journal B}, 26:521--529, 2002.

\bibitem{10.1007/978-3-319-61176-1_2}
Dieudonne Mulamba and Indrajit Ray.
\newblock Resilient reference monitor for distributed access control via moving
  target defense.
\newblock In Giovanni Livraga and Sencun Zhu, editors, {\em Data and
  Applications Security and Privacy XXXI}, pages 20--40, 2017.

\bibitem{neuhaus2007predicting}
Stephan Neuhaus, Thomas Zimmermann, Christian Holler, and Andreas Zeller.
\newblock Predicting vulnerable software components.
\newblock In {\em Proceedings of the 2007 {ACM} Conference on Computer and
  Communications Security, {CCS} 2007, Alexandria, Virginia, USA}, pages
  529--540, 2007.

\bibitem{NewmanReview2003}
M.~Newman.
\newblock The structure and function of complex networks.
\newblock {\em SIAM Review}, 45:167, 2003.

\bibitem{NSAHardProblemList}
David Nicol, Bill Sanders, Jonathan Katz, Bill Scherlis, Tudor Dumitra, Laurie
  Williams, and Munindar~P. Singh.
\newblock The science of security 5 hard problems (august 2015).
\newblock \url{http://cps-vo.org/node/21590}.

\bibitem{DBLP:journals/tdsc/NicolST04}
David~M. Nicol, William~H. Sanders, and Kishor~S. Trivedi.
\newblock Model-based evaluation: From dependability to security.
\newblock {\em {IEEE} Trans. Dependable Sec. Comput.}, 1(1):48--65, 2004.

\bibitem{Noel2017}
Steven Noel and Sushil Jajodia.
\newblock {\em A Suite of Metrics for Network Attack Graph Analytics}, pages
  141--176.
\newblock Springer International Publishing, Cham, 2017.

\bibitem{MIT-MTD-Survey-2013}
H.~Okhravi, M.~Rabe, T.~Mayberry, W.~Leonard, T.~Hobson, D.~Bigelow, and
  W.~Streilein.
\newblock Survey of cyber moving targets (mit lincoln lab technical report),
  2013.

\bibitem{XuIntrust09-keysecurity}
T.~Paul Parker and Shouhuai Xu.
\newblock A method for safekeeping cryptographic keys from memory disclosure
  attacks.
\newblock In {\em First International Conference on Trusted Systems
  (INTRUST'2009)}, pages 39--59, 2009.

\bibitem{Pastor2001}
R.~Pastor-Satorras and A.~Vespignani.
\newblock Epidemic dynamics and endemic states in complex networks.
\newblock {\em Physical Review E}, 63:066117, 2001.

\bibitem{Pastor2002a}
R.~Pastor-Satorras and A.~Vespignani.
\newblock Epidemic dynamics in finite size scale-free networks.
\newblock {\em Physical Review E}, 65:035108, 2002.

\bibitem{Pendleton16}
Marcus Pendleton, Richard Garcia-Lebron, Jin-Hee Cho, and Shouhuai Xu.
\newblock A survey on systems security metrics.
\newblock {\em ACM Comput. Surv.}, 49(4):62:1--62:35, December 2016.

\bibitem{XuMarkerPointProcess2017}
Chen Peng, Maochao Xu, Shouhuai Xu, and Taizhong Hu.
\newblock Modeling and predicting extreme cyber attack rates via marked point
  processes.
\newblock {\em Journal of Applied Statistics}, 44(14):2534--2563, 2017.

\bibitem{XuJournalAppliedStatistics2018}
Chen Peng, Maochao Xu, Shouhuai Xu, and Taizhong Hu.
\newblock Modeling multivariate cybersecurity risks.
\newblock {\em Journal of Applied Statistics}, 0(0):1--23, 2018.

\bibitem{Perdisci:2012:VTF:2420950.2420999}
Roberto Perdisci and ManChon U.
\newblock Vamo: Towards a fully automated malware clustering validity analysis.
\newblock In {\em Proceedings of the 28th Annual Computer Security Applications
  Conference}, ACSAC '12, pages 329--338, 2012.

\bibitem{Pfleeger:2009:UCM:1591880.1592054}
Shari~Lawrence Pfleeger.
\newblock Useful cybersecurity metrics.
\newblock {\em IT Professional}, 11(3):38--45, 2009.

\bibitem{5432146}
S.L. Pfleeger and R.K. Cunningham.
\newblock Why measuring security is hard.
\newblock {\em Security Privacy, IEEE}, 8(4):46--54, July 2010.

\bibitem{Phillips:1998:GSN:310889.310919}
Cynthia Phillips and Laura~Painton Swiler.
\newblock A graph-based system for network-vulnerability analysis.
\newblock In {\em Proc. 1998 Workshop on New Security Paradigms}, NSPW '98,
  pages 71--79, 1998.

\bibitem{ProvosUsenixSecurity04}
N.~Provos.
\newblock A virtual honeypot framework.
\newblock In {\em USENIX Security Symposium}, pages 1--14, 2004.

\bibitem{ProvosHotbot07}
N.~Provos, D.~McNamee, P.~Mavrommatis, K.~Wang, and N.~Modadugu.
\newblock The ghost in the browser analysis of web-based malware.
\newblock In {\em Proceedings of the First Workshop on Hot Topics in
  Understanding Botnets (HotBots'07)}, 2007.

\bibitem{DBLP:conf/ccs/RahmanAB14}
Mohammad~Ashiqur Rahman, Ehab Al{-}Shaer, and Rakesh~B. Bobba.
\newblock Moving target defense for hardening the security of the power system
  state estimation.
\newblock In {\em Proceedings of the First {ACM} Workshop on Moving Target
  Defense, {MTD} '14}, pages 59--68, 2014.

\bibitem{8017389}
A.~Ramos, M.~Lazar, R.~H. Filho, and J.~J. P.~C. Rodrigues.
\newblock Model-based quantitative network security metrics: A survey.
\newblock {\em IEEE Communications Surveys Tutorials}, 19(4):2704--2734, 2017.

\bibitem{Ritchey:2000:UMC:882494.884423}
Ronald~W. Ritchey and Paul Ammann.
\newblock Using model checking to analyze network vulnerabilities.
\newblock In {\em Proc. IEEE Symposium on Security and Privacy}, pages
  156--165, 2000.

\bibitem{DBLP:conf/hotsos/RoqueBD16}
Antonio Roque, Kevin~B. Bush, and Christopher Degni.
\newblock Security is about control: insights from cybernetics.
\newblock In {\em Proceedings of the Symposium and Bootcamp on the Science of
  Security, Pittsburgh, PA, USA, April 19-21, 2016}, pages 17--24, 2016.

\bibitem{Schneider11_BlueprintForScienceOfCybersecurity}
Fred Schneider.
\newblock Blueprint for a science of cybersecurity.
\newblock Technical report, Cornell University, May 2011.
\newblock Also to appear in The Next Wave.

\bibitem{NITRD}
National Science and Technology Council.
\newblock Trustworthy cyberspace: Strategic plan for the federal cybersecurity
  research and development program.
\newblock
  \url{https://www.nitrd.gov/SUBCOMMITTEE/csia/Fed_Cybersecurity_RD_Strategic_Plan_2011.pdf},
  2011.

\bibitem{XuPRE2011}
Y.~Shang, W.~Luo, and S.~Xu.
\newblock $l$-hop percolation on networks with arbitrary degree distributions
  and its applications.
\newblock {\em Phys. Rev. E}, 84:031113, Sep 2011.

\bibitem{JhaOkland02}
O.~Sheyner, J.~Haines, S.~Jha, R.~Lippmann, and J.~Wing.
\newblock Automated generation and analysis of attack graphs.
\newblock In {\em IEEE Symposium on Security and Privacy}, pages 273--284,
  2002.

\bibitem{shin2011evaluating}
Yonghee Shin, Andrew Meneely, Laurie Williams, and Jason~A Osborne.
\newblock Evaluating complexity, code churn, and developer activity metrics as
  indicators of software vulnerabilities.
\newblock {\em IEEE Transactions on Software Engineering}, 37(6):772--787,
  2011.

\bibitem{DBLP:conf/nspw/SpringMP17}
Jonathan~M. Spring, Tyler Moore, and David~J. Pym.
\newblock Practicing a science of security: {A} philosophy of science
  perspective.
\newblock In {\em Proceedings of the 2017 New Security Paradigms Workshop,
  {NSPW} 2017}, pages 1--18, 2017.

\bibitem{DBLP:journals/corr/abs-1802-03802}
Caroline Trippel, Daniel Lustig, and Margaret Martonosi.
\newblock Meltdownprime and spectreprime: Automatically-synthesized attacks
  exploiting invalidation-based coherence protocols.
\newblock {\em CoRR}, abs/1802.03802, 2018.

\bibitem{XuPhysicaA2017}
Adam Tyra, Jingtao Li, Yilun Shang, Shuo Jiang, Yanjun Zhao, and Shouhuai Xu.
\newblock Robustness of non-interdependent and interdependent networks against
  dependent and adaptive attacks.
\newblock {\em Physica A: Statistical Mechanics and its Applications}, 482:713
  -- 727, 2017.

\bibitem{VanMieghemIEEEACMTON09}
Piet Van~Mieghem, Jasmina Omic, and Robert Kooij.
\newblock Virus spread in networks.
\newblock {\em IEEE/ACM Trans. Netw.}, 17(1):1--14, February 2009.

\bibitem{WangSRDS03}
Y.~Wang, D.~Chakrabarti, C.~Wang, and C.~Faloutsos.
\newblock Epidemic spreading in real networks: An eigenvalue viewpoint.
\newblock In {\em Proc. of the 22nd IEEE Symposium on Reliable Distributed
  Systems (SRDS'03)}, pages 25--34, 2003.

\bibitem{XuCODASPY13}
Li~Xu, Zhenxin Zhan, Shouhuai Xu, and Keying Ye.
\newblock Cross-layer detection of malicious websites.
\newblock In {\em Third {ACM} Conference on Data and Application Security and
  Privacy (ACM CODASPY'13)}, pages 141--152, 2013.

\bibitem{XuCNS2014}
Li~Xu, Zhenxin Zhan, Shouhuai Xu, and Keying Ye.
\newblock An evasion and counter-evasion study in malicious websites detection.
\newblock In {\em IEEE Conference on Communications and Network Security
  (CNS'14)}, pages 141--152, 2013.

\bibitem{XuTIFSDataBreach2018}
M.~Xu, K.~M. Schweitzer, R.~M. Bateman, and S.~Xu.
\newblock Modeling and predicting cyber hacking breaches.
\newblock {\em IEEE Transactions on Information Forensics and Security},
  13(11):2856--2871, Nov 2018.

\bibitem{XuInternetMath2015Dependence}
Maochao Xu, Gaofeng Da, and Shouhuai Xu.
\newblock Cyber epidemic models with dependences.
\newblock {\em Internet Mathematics}, 11(1):62--92, 2015.

\bibitem{XuVineCopula2015}
Maochao Xu, Lei Hua, and Shouhuai Xu.
\newblock A vine copula model for predicting the effectiveness of cyber defense
  early-warning.
\newblock {\em Technometrics}, 59(4):508--520, 2017.

\bibitem{XuInternetMath2012}
Maochao Xu and Shouhuai Xu.
\newblock An extended stochastic model for quantitative security analysis of
  networked systems.
\newblock {\em Internet Mathematics}, 8(3):288--320, 2012.

\bibitem{XuAsiaCCS08}
S.~Xu, X.~Li, and P.~Parker.
\newblock Exploiting social networks for threshold signing: Attack-resilience
  vs. availability.
\newblock In {\em ACM Symposium on Information, Computer and Communications
  Security (ASIACCS'08)}, pages 325--336, 2008.

\bibitem{XuIEEETIFS2011}
S.~Xu, X.~Li, T.~Parker, and X.~Wang.
\newblock Exploiting trust-based social networks for distributed protection of
  sensitive data.
\newblock {\em IEEE Transactions on Information Forensics and Security},
  6(1):39--52, 2011.

\bibitem{XuCD}
Shouhuai Xu.
\newblock Cybersecurity dynamics publications.
\newblock \url{http://www.cs.utsa.edu/~shxu/socs/}.

\bibitem{DBLP:journals/jcs/Xu07}
Shouhuai Xu.
\newblock On the security of group communication schemes.
\newblock {\em Journal of Computer Security}, 15(1):129--169, 2007.

\bibitem{XuCollaborateCom08}
Shouhuai Xu.
\newblock Collaborative attack vs. collaborative defense.
\newblock In {\em Collaborative Computing: Networking, Applications and
  Worksharing, 4th International Conference (CollaborateCom'2008)}, pages
  217--228, 2008.

\bibitem{XuCybersecurityDynamicsHotSoS2014}
Shouhuai Xu.
\newblock Cybersecurity dynamics.
\newblock In {\em Proc. Symposium and Bootcamp on the Science of Security
  (HotSoS'14)}, pages 14:1--14:2, 2014.

\bibitem{XuEmergentBehaviorHotSoS2014}
Shouhuai Xu.
\newblock Emergent behavior in cybersecurity.
\newblock In {\em Proceedings of the 2014 Symposium and Bootcamp on the Science
  of Security (HotSoS'14)}, pages 13:1--13:2, 2014.

\bibitem{XuInternetMath2015ACD}
Shouhuai Xu, Wenlian Lu, and Hualun Li.
\newblock A stochastic model of active cyber defense dynamics.
\newblock {\em Internet Mathematics}, 11(1):23--61, 2015.

\bibitem{XuTAAS2012}
Shouhuai Xu, Wenlian Lu, and Li~Xu.
\newblock Push- and pull-based epidemic spreading in arbitrary networks:
  Thresholds and deeper insights.
\newblock {\em ACM Transactions on Autonomous and Adaptive Systems (ACM TAAS)},
  7(3):32:1--32:26, 2012.

\bibitem{XuTAAS2014}
Shouhuai Xu, Wenlian Lu, Li~Xu, and Zhenxin Zhan.
\newblock Adaptive epidemic dynamics in networks: Thresholds and control.
\newblock {\em ACM Transactions on Autonomous and Adaptive Systems (ACM TAAS)},
  8(4):19, 2014.

\bibitem{XuTDSC2012}
Shouhuai Xu, Wenlian Lu, and Zhenxin Zhan.
\newblock A stochastic model of multivirus dynamics.
\newblock {\em IEEE Transactions on Dependable and Secure Computing},
  9(1):30--45, 2012.

\bibitem{XuFC09}
Shouhuai Xu and Moti Yung.
\newblock Expecting the unexpected: Towards robust credential infrastructure.
\newblock In {\em Financial Cryptography and Data Security, 13th International
  Conference (FC'09)}, pages 201--221, 2009.

\bibitem{yamaguchi2011vulnerability}
Fabian Yamaguchi, Felix~"FX" Lindner, and Konrad Rieck.
\newblock Vulnerability extrapolation: Assisted discovery of vulnerabilities
  using machine learning.
\newblock In {\em 5th {USENIX} Workshop on Offensive Technologies, WOOT'11,
  August 8, 2011, San Francisco, CA, USA, Proceedings}, pages 118--127, 2011.

\bibitem{yamaguchi2012generalized}
Fabian Yamaguchi, Markus Lottmann, and Konrad Rieck.
\newblock Generalized vulnerability extrapolation using abstract syntax trees.
\newblock In {\em 28th Annual Computer Security Applications Conference,
  {ACSAC} 2012, Orlando, FL, USA}, pages 359--368, 2012.

\bibitem{yamaguchi2013chucky}
Fabian Yamaguchi, Christian Wressnegger, Hugo Gascon, and Konrad Rieck.
\newblock Chucky: Exposing missing checks in source code for vulnerability
  discovery.
\newblock In {\em 2013 {ACM} {SIGSAC} Conference on Computer and Communications
  Security, CCS'13, Berlin, Germany}, pages 499--510, 2013.

\bibitem{XuTIFS2013}
Zhenxin Zhan, Maochao Xu, and Shouhuai Xu.
\newblock Characterizing honeypot-captured cyber attacks: Statistical framework
  and case study.
\newblock {\em IEEE Transactions on Information Forensics and Security},
  8(11):1775--1789, 2013.

\bibitem{XuInTrust2014}
Zhenxin Zhan, Maochao Xu, and Shouhuai Xu.
\newblock A characterization of cybersecurity posture from network telescope
  data.
\newblock In {\em Proc. of the 6th International Conference on Trustworthy
  Systems (InTrust'14)}, pages 105--126, 2014.

\bibitem{XuTIFS2015}
Zhenxin Zhan, Maochao Xu, and Shouhuai Xu.
\newblock Predicting cyber attack rates with extreme values.
\newblock {\em {IEEE} Transactions on Information Forensics and Security},
  10(8):1666--1677, 2015.

\bibitem{XieNSDI09}
Y.~Zhao, Y.~Xie, F.~Yu, Q.~Ke, Y.~Yu, Y.~Chen, and E.~Gillum.
\newblock Botgraph: large scale spamming botnet detection.
\newblock In {\em Proc. NSDI'09}, pages 321--334, 2009.

\bibitem{XuCODASPY2011-POR}
Qingji Zheng and Shouhuai Xu.
\newblock Fair and dynamic proofs of retrievability.
\newblock In {\em First {ACM} Conference on Data and Application Security and
  Privacy, (CODASPY'2011)}, pages 237--248, 2011.

\bibitem{XuCODASPY2012}
Qingji Zheng and Shouhuai Xu.
\newblock Secure and efficient proof of storage with deduplication.
\newblock In {\em Second ACM Conference on Data and Application Security and
  Privacy (CODASPY'2012)}, pages 1--12, 2012.

\bibitem{XuIC2E2015}
Qingji Zheng and Shouhuai Xu.
\newblock Verifiable delegated set intersection operations on outsourced
  encrypted data.
\newblock In {\em 2015 {IEEE} International Conference on Cloud Engineering,
  {IC2E} 2015}, pages 175--184, 2015.

\bibitem{XuCCSW12}
Qingji Zheng, Shouhuai Xu, and Giuseppe Ateniese.
\newblock Efficient query integrity for outsourced dynamic databases.
\newblock In {\em Proceedings of the 2012 {ACM} Workshop on Cloud computing
  security, {CCSW} 2012, Raleigh, NC, USA, October 19, 2012.}, pages 71--82,
  2012.

\bibitem{XuInfocom2014}
Qingji Zheng, Shouhuai Xu, and Giuseppe Ateniese.
\newblock {VABKS:} verifiable attribute-based keyword search over outsourced
  encrypted data.
\newblock In {\em Proc. 2014 {IEEE} Conference on Computer Communications
  (INFOCOM'2014)}, pages 522--530, 2014.

\bibitem{XuTNSE2018}
R.~Zheng, W.~Lu, and S.~Xu.
\newblock Preventive and reactive cyber defense dynamics is globally stable.
\newblock {\em IEEE Transactions on Network Science and Engineering}, pages
  1--1, 2017.

\bibitem{XuHotSoS2015}
Ren Zheng, Wenlian Lu, and Shouhuai Xu.
\newblock Active cyber defense dynamics exhibiting rich phenomena.
\newblock In {\em Proc. 2015 Symposium and Bootcamp on the Science of Security
  (HotSoS'15)}, pages 2:1--2:12, 2015.

\bibitem{XUJCS06}
S.~Zhu, S.~Setia, S.~Xu, and S.~Jajodia.
\newblock Gkmpan: An efficient group rekeying scheme for secure multicast in
  ad-hoc networks.
\newblock {\em Journal of Computer Security}, 14(4):301--325, 2006.

\end{thebibliography}

\end{document}